\documentclass[review]{elsarticle}
\usepackage{hyperref}

\usepackage[utf8]{inputenc}  
\usepackage{listings}
\usepackage{color}
\usepackage{amssymb}  
\usepackage{amsmath}  
\usepackage{bm}  
\usepackage{subcaption}  

\usepackage{array}
\newcolumntype{L}[1]{>{\raggedright\let\newline\\\arraybackslash\hspace{0pt}}m{#1}}
\newcolumntype{C}[1]{>{\centering\let\newline\\\arraybackslash\hspace{0pt}}m{#1}}
\newcolumntype{R}[1]{>{\raggedleft\let\newline\\\arraybackslash\hspace{0pt}}m{#1}}

\usepackage{siunitx}  
\makeatletter
\providecommand{\doi}[1]{%
	\begingroup
	\let\bibinfo\@secondoftwo
	\urlstyle{rm}%
	\href{http://dx.doi.org/#1}{%
		doi:\discretionary{}{}{}%
		\nolinkurl{#1}%
	}%
	\endgroup
}
\makeatother

\definecolor{cadmiumgreen}{rgb}{0.0, 0.42, 0.24}

\lstdefinestyle{floatcode}{float=tp,floatplacement=tbp
}

\newcommand\changed[1]{#1}

\newcommand\descriptor{\mathcal{D}}
\newcommand\comparator{\mathcal{C}}
\newcommand\ranker{R}
\newcommand\rank{\tau}
\newcommand\rankof[1]{\rank_{#1}}
\newcommand\rankposition[2]{\rho_{#1}(#2)}
\newcommand\ranks{\mathcal{T}}
\newcommand\ranksof[1]{\mathcal{T}_{#1}}
\newcommand\scoresymbol{\varsigma}
\newcommand\score[2]{\scoresymbol(#1,#2)}
\newcommand\scorein[3]{\varsigma_{#1}(#2,#3)}

\journal{Journal of Information Processing and Management}

\biboptions{authoryear}
\renewcommand{\cite}[1]{\citep{#1}}  

\begin{document}

\sloppy  

\begin{frontmatter}

\title{Unsupervised Graph-based Rank Aggregation for Improved Retrieval}

\author[address1]{Icaro Cavalcante Dourado\corref{correspondingauthor}}
\ead{icaro.dourado@ic.unicamp.br}

\author[address2]{Daniel Carlos Guimarães Pedronette}
\ead{daniel@rc.unesp.br}

\author[address1]{Ricardo da Silva Torres}
\ead{rtorres@ic.unicamp.br}

\cortext[correspondingauthor]{Corresponding author.}

\address[address1]{Institute of Computing, University of Campinas (UNICAMP), Campinas, Brazil}
\address[address2]{Department of Statistics, Applied Mathematics and Computing, São Paulo State University (UNESP), Rio Claro, Brazil}

\begin{abstract}
This paper presents a robust and comprehensive graph-based rank aggregation approach, used to combine results of isolated ranker models in retrieval tasks.
The method follows an unsupervised scheme, which is independent of how the isolated ranks are formulated.
Our approach is able to combine arbitrary models, defined in terms of different ranking criteria, such as those based on textual, image or hybrid content representations.

\changed{We reformulate the ad-hoc retrieval problem as a document retrieval based on {\em fusion graphs},}
which we propose as a new unified representation model capable of merging multiple ranks and expressing inter-relationships of retrieval results automatically. By doing so, we claim that the retrieval system can benefit from learning the manifold structure of datasets, thus leading to more effective results.
Another contribution is that our graph-based aggregation formulation, unlike existing approaches, allows for encapsulating contextual information encoded from multiple ranks\changed{, which can be directly used for ranking, without further computations and post-processing steps over the graphs.}
Based on the graphs, a novel similarity retrieval score is formulated using an efficient computation of minimum common subgraphs.
Finally, another benefit over existing approaches is the absence of hyperparameters.

A comprehensive experimental evaluation was conducted considering diverse well-known public datasets, composed of textual, image, and multimodal documents.
Performed experiments demonstrate that our method reaches top performance, yielding better effectiveness scores than state-of-the-art baseline methods and promoting large gains over the rankers being fused,
\changed{thus demonstrating the successful capability of the proposal in representing queries based on a unified graph-based model of rank fusions.}
\end{abstract}


\end{frontmatter}


\section{Introduction}

The increasing demand of effective and efficient retrieval methods, due to the huge growth of the volume \changed{and diversity} of available data, has encouraged the creation of sophisticated feature extraction algorithms.
\changed{These algorithms are important as they are the basis of subsequent generalization and learning models, commonly used in several domains, such as search and classification tasks.}
The proposal of description approaches for images, texts, and multimedia data has advanced in the last decades\changed{, leading to more discriminative and effective models}.
However, the choice of the most suitable technique often depends on the circumstances (e.g., application or dataset) in which they are used. In fact, an active research venue relies on exploiting their complementary view, by aggregation, aiming to improve the effectiveness of complex services, such as search, classification, or recommendation.

Rank aggregation techniques are important in many applications, such as meta-search, document filtering, recommendation systems, social choice, etc. Unsupervised and supervised rank aggregation methods have been proposed in order to combine results from different rankers and promote more effective retrieval results.
Although supervised methods have the potential to produce better fusions, in practice they demand more computational cost, and require training data that may be either unavailable or expensive to obtain.
\changed{A crowd paradigm, aimed at obtaining labeled training data through voluntary or paid collaborative work, can mitigate the lack of training data. However, this labeling task can still be a time-consuming, expensive, and unfeasible process; or even introduce bias to data.}

Several different strategies have been exploited by rank aggregation methods, mainly based on available information provided by retrieval scores~\cite{fox:1994:combination} or positions in ranks~\cite{borda:1784:bordaCount,cormack2:2009:RRF}.
Another common approach is based on Markov Chain, where retrieved objects are represented in the various ranks as vertices in a graph, and transition probabilities from vertex to vertex are defined in terms the relative positions of the items in the various ranks~\cite{PaperSimilarItems_Sculley2007,RankAggregForWeb_WWW01}.
In fact, graphs have been proved to be a powerful tool for modeling the relationships among data objects in recent rank aggregation approaches~\cite{zhang:2015:queryRankFusion,pedronette:2017:reciprocalGraph}.

In this paper, we propose an unsupervised graph-based rank aggregation method, agnostic of the rankers being fused, and targeted for general applicability, such as image, textual, or even multimodal retrieval tasks.
\changed{We reformulate the ad-hoc retrieval problem as a document retrieval based on fusion graphs,}
which we propose as a new unified representation model capable of merging multiple ranks and express inter-relationships of retrieval results automatically. By doing so, our main research objective is to investigate our hypothesis that the retrieval system can benefit from learning the manifold structure of datasets, thus promoting more effective results.

\changed{As we model and retrieve objects by multiple rankers, the main application of our method is ad-hoc retrieval, also called content-based retrieval, in which images, documents or even multimodal objects are used as queries in a retrieval system. This application can be adopted in several applications, such as digital libraries, social media, and service providers. Content-based image retrieval (CBIR), for instance, is a promising field for application.}

Another research objective consists in investigating the impact of different ranker selection criteria for fusion, which take into account the rankers' effectiveness and their correlation.

Different from most related rank aggregation approaches, the proposed method does not require free parameters, such as neighborhood size definition for the graph construction.
The proposed method is also innovative with regard to the definition of the fused retrieval score. While other related graph-based approaches exploit the graph through operations on transition matrices~\cite{zhang:2015:queryRankFusion} or specific similarity measures~\cite{pedronette:2017:reciprocalGraph}, our approach derives a new retrieval score directly based on the graph structure, considering the minimum common subgraph of two objects' graphs. In summary, our fusion method relies on exploiting contextual information obtained from the direct comparison of objects based on their neighbors, which are defined in terms of the ranks associated with multiple ranking criteria.

\changed{The contributions of the work are:
\begin{enumerate}
\item The proposal of a novel graph-based rank aggregation model,
\begin{itemize}
    \item which is unsupervised, does not require tuning of hyperparameters, and yields top performance compared to state-of-the-art baselines and large gains over the rankers being fused;
    \item which is agnostic about the ranks, such as how they are generated, their weighting functions, or whether they are based on distance or similarity scores;
    \item which is flexible as its components, the fusion graph extraction and the graph-based retrieval, are independent, both being capable of adaptation or further improvement.
\end{itemize}
\item The proposal of {\em fusion graphs}, a graph representation, which is capable of merging multiple ranks and expressing inter-relationships of retrieval results automatically. The proposed representation intrinsically supports multimodal objects, meaning that it can be applied over ranks defined according to different data types at same time;
\item Unlike existing approaches, a straightforward ranking procedure is proposed, for the fusion representation. The method does not require optimizations or additional processing steps;
\item A novel similarity score is formulated, based on the fusion graphs, using an efficient computation of minimum common subgraphs.
\end{enumerate}}

The remainder of this paper is organized as follows. Section~\ref{secRW} discusses related work and Section~\ref{secRetrModel} formally describes the retrieval model considered. Section~\ref{propMethod} presents the proposed method and Section~\ref{secExpEval} describes conducted experiments.
Finally, Section~\ref{secConc} concludes the paper and provides possible future research directions.

\section{Related Work}
\label{secRW}

\changed{The Kemeny ranking problem aims to get a consensus rank (or median rank) that best represents a given set of ranks, i.e. an optimal permutation that best summarizes them.
Its general case, known as rank aggregation problem (RAP), targets any kind of rank, complete or incomplete, and with or without ties. A rank is incomplete if it does not contain all the items. A tie in a rank, in turn, refers to the presence of equally preferred items.}

\changed{There is a family of initiatives that address rank aggregation from a theoretical perspective of optimal or sub-optimal aggregations.
\citet{Aledo:2016:ExtensionSetsFlexibleRank} defined an {\em extension set} of a rank as the set of permutations that are {\em compatible} (of equivalent importance) with the given rank, and then proposed a solution for RAP that allows any ranks to be aggregated, based on extension sets.
\citet{Amodio:2016:accurateMedianRanking} proposed a heuristic algorithm for RAP that finds one of the existing optimal median ranks in less computational time than more expensive branch-and-bound methods.
\citet{DAmbrosio:2017:EvolutionMedianRanking} proposed an evolutionary heuristic for RAP, called differential evolution algorithm, that is able to deal with a large number of items in reasonable time, when compared to branch-and-bound and other heuristics.
Similarly, \citet{Aledo:2018:rankAggEvolution} presented evolutionary approaches, and studied the effect of mutation operators, initialization methods, and generation of descendants.
Both evolutionary approaches surpassed previous greedy methods.}

\changed{Anyway, RAP is a NP-hard problem for more than three input ranks~\cite{RankAggregForWeb_WWW01}.
In practice, rank aggregation can be seen as the task of finding a good permutation of retrieved objects obtained from different input ranks. In this case, rank aggregation methods compose inexact solutions that intend to promote better results than the isolated input ranks. Note that these RAP-based theoretical works have not been explored for retrieval tasks either.}

Related to the rank aggregation task, \textit{re-ranking} refers to a prior family of methods that also intend to promote better results, but do not explore the inter-relationships between the ranks from the response objects. Re-ranking approaches are feature-based~\cite{Hubert:2018:TournaRank} or rank-based~\cite{bai:2016:sparse}. In this sense, the exploitation of inter-relationships between ranks is a potential advantage for the rank aggregation methods by definition. Besides, the main advantage of ranked-based approaches for improved retrieval, over feature-based approaches, is that while digital objects are typically modeled in high dimensional spaces, they often live in a much lower-dimensional intrinsic manifold space~\cite{Zhao:2018:metricsRelations}. For this reason, ranked-based approaches can be more efficient while assuming less input data.

\changed{Supervised rank aggregation methods are intended to infer fusion formulations automatically from training data, by exploiting labeled information and ground-truth relevance to maximize the effectiveness of a new ranker. Supervised rank aggregation methods belong to the Learning-to-Ranking (L2R) field, which refers to a broader family of supervised methods for ranking. As a drawback, the availability of training data is not always possible or feasible, and supervised techniques demand more computational cost.}

\changed{\citet{Kaur:2017:rankAggMetasearchGenetProg} proposed a metaheuristic approach, based on genetic algorithm (GA), which optimizes the rank aggregation problem for search engines.
The GA approach applies an optimization process over distance measures to minimize the distances for various aggregated ranks to generate a final aggregated rank.
\citet{Mourao:2018:LowComplexityL2R} proposed Learning to Fuse (L2F), a L2R algorithm of presumably lower complexity than other more costly L2R models but with competitive retrieval results to them.
Their solution mitigates the final complexity by analyzing and discarding ranks of minor improvements to final rank, during its learning process, thus trading precision for complexity.
In our opinion, such supervised rank aggregation models are still either too complex, data-dependent or costly to scenarios in which unsupervised models can be satisfactory.
Here we focus on unsupervised methods for rank aggregation.}

\changed{Less frequently, some initiatives proposed aggregation methods that work upon both object features and ranks.
This is a promising approach, but also demands raw data, which may not be available in practical situations.
\citet{Bhowmik:2017:letorUnsupRankAgg} proposed a hybrid unsupervised rank aggregation method that is based on both object attributes and ranks, as an augmented solution.
Furthermore, their evaluation considered only a few classic baselines (up to 2001) that did not explore both aspects either.}

\changed{Rank aggregation methods can be  also classified as either order-based or score-based.}
Order-based methods use the relative order among the retrieved objects to aggregate the ranks.
Score-based methods also use the scores associated with each retrieved object from different ranks as input.

BordaCount~\cite{borda:1784:bordaCount} is an order-based method
that computes a new score of each retrieved object based on the disparity between its positions on the ranks with respect to the their sizes.
Reciprocal Rank Fusion (RRF)~\cite{cormack2:2009:RRF}, by contrast, is an order-based method that assigns scores to retrieved objects using a formulation that more emphatically penalizes lower-lanker results in favor of highly ranked results.

Median Rank Aggregation (MRA)~\cite{fagin:2003:MRA} is an order-based method.
It traverses the ranks counting the number of occurrences of the retrieved objects.
The first object that occurs in more than half of the ranks is taken as the first object of the final rank.
Then, the second object that occurs in more than half of the ranks is taken as the second, and so on.

Six score-based methods were proposed by~\citet{fox:1994:combination}: CombSUM, CombMAX, CombMIN, CombMED, CombMNZ, and CombANZ, based on distinct priors. 
For these methods, each rank must be previously normalized with respect to its scores.
Related to these methods, RLSim~\cite{pedronette:2013:rlsim} is a score-based technique, inspired by Naive Bayes classifier, that assigns the final score of an object as the product of its scores in each rank.

Condorcet is a voting method, based on the Condorcet criterion.
This criterion defines that the winner of the election is the candidate that beats the other candidates in pairwise comparisons.
Let the distance between two ranks be the number of pairs whose objects are ranked reversely. The Condorcet winner is the one that minimizes the total distance.
The Condorcet method produces a ranking of all candidates from the first to the last place. The Condorcet winner comes first and the Condorcet loser comes last.

Some graph-based approaches for rank fusion were proposed based on Markov Chains, where retrieved objects are represented in the various ranks as vertices in a graph, with transition probabilities from vertex to vertex defined by the relative rankings of the items in the various ranks~\cite{PaperSimilarItems_Sculley2007,RankAggregForWeb_WWW01}.

\citet{zhang:2015:queryRankFusion} proposed a graph-based rank aggregation method, named here as QueryRankFusion, that explores the notion of reciprocal references.
It analyzes the $k$-reciprocal neighborhoods for building a graph for each rank, and requires the computation of the Jaccard measure for assigning weights to edges.
\changed{Graphs are later fused into a global graph. Then, it relies on a ranking step using two possible solvers, either based on the PageRank algorithm that computes a transition matrix over the edges or by a greedy algorithm that finds subgraphs of maximum local density.
This method depends on the adjustment of three hyperparameters: the number $k$ of neighbors to analyze; the solver algorithm for the ranking step; and the number of iterations in the ranking step.}
This method yields effective results, but it was validated only for image retrieval tasks.

Similarly, \citet{pedronette:2016:RkGraph} proposed RkGraph, a graph-based aggregation approach for distance learning in shape retrieval tasks, which merges graphs defined upon multiple ranks and composes a collection graph.

\citet{pedronette:2016:CorGraph} proposed CorGraph, a learning method based on a correlation graph, which defines the graph connectivity using different levels of correlation measures and exploits strongly connected components. \citet{pedronette:2017:reciprocalGraph} continued their work proposing a simpler graph-based method, hereby called RecKNNGraphCCs, as in~\citet{zhang:2015:queryRankFusion}, but with less intermediate steps and less hyperparameters. In their method, they rely on connected components in the step of generating ranks. A pre-processing step of re-ranking and normalization is performed to improve the ranks before the use of the rank aggregation scheme.
This method is affected by two hyperparameters: the number of iterations and the number of neighbors to analyze.
This method was also validated only in image retrieval problems.

Existing graph-based methods are mostly targeted at modelling the whole collection of objects as a graph, from which the ranks can be derived.
Different from these works, we model one fusion graph per object, and redefine the object retrieval system by means of fusion graphs.
Our graph definition not only encapsulates the information from ranks of a certain query, but also incorporates information regarding inter-relationships between the results from the ranks.

\changed{Our approach presents theoretical and practical advances and implications.
A theoretical implication is that ranks can be directly used for fusion, thus promoting a unified representation. From this representation, called fusion graph, we derive a straightforward ranking procedure, without further transformations and optimizations. That corresponds to the second theoretical implication.
Finally, a practical implication from our approach is its advantage that the fusion graph extraction and the graph-based retrieval are independent, both being capable of adaptation or further improvement.
In addition, our solution does not require time-consuming tuning of hyperparameters.}


\section{Retrieval Model}
\label{secRetrModel}

Our approach for ranking is based upon the following definitions.
Let ${S = \{ s_1, s_2, \ldots, s_n\}}$ be a collection of $n$ \textit{samples}, where $n$ is the collection size.
A sample can be a document, an image, a video, or even a hybrid (called multimodal) object.
Each sample $s$ from $S$ is characterized (or described) by \textit{descriptors}. Each descriptor, $\descriptor$, has its own assumption, pros and cons, and represents a specific point of view with respect to the samples. For this reason, it is common to use multiple descriptors to characterize a collection. A descriptor is used to assign (extract from the object) to $s$, a vector, a graph, or any data structure $\epsilon(s)$. The purpose of $\epsilon(s)$ is to allow the comparison of objects, supporting the creation of services, such as search or recommendation.
Comparisons are performed by means of a \textit{comparator}, $\comparator$, applied over a tuple $(\epsilon(s_i),\epsilon(s_j))$, that produces a \textit{score} $\scoresymbol$ of codomain $\mathbb{R^+}$ (e.g., the cosine similarity and the Euclidean distance) in the sense that either similarity or dissimilarity functions can be used. 
We employ a special procedure to convert dissimilarity into similarity scores, so that different kinds of ranks can be combined. We present our standardization procedure in Section~\ref{sec:rankNormalization}.

A query sample $q$, or just \textit{query}, follows the same definition of a sample, but refers to the input object in the context of a search: a search brings \textit{response items} (samples) from the \textit{response set} ($S$) according to a relevance criteria. 
We refer to a \textit{ranker} as a tuple $\ranker = (\descriptor,\comparator)$, which is employed to compute a rank, $\rank$, for $q$. \changed{A ranker may be seen as a simplified \textit{retrieval model}~\cite{baeza:1999:modernIR}.} We denote by $\rankof{q}$ the rank for a certain input query $q$. A rank is a permutation of $S_L \subseteq S$, where $L \ll n$ in general, such that $\rankof{q}$ provides the most similar -- or equivalently the least dissimilar -- samples, to $q$, from $S$, in order. $L$ is used as a cut-off parameter. $\rankposition{\rankof{q}}{x}$ refers to the position of $x$ in $\rankof{q}$.

Given a certain ranker $\ranker$, 
we may refer to $\score{s_i}{s_j}$ as meaning $\score{\epsilon(s_i)}{\epsilon(s_j)}$, for simplicity.
$s_i$ occurs closer to the first positions of $\rankof{q}$ than $s_j$ if $\score{q}{s_i} \leq \score{q}{s_j}$.
Along with one rank $\rankof{q}$ -- a sequence of objects $s_i$ -- we also hold the scores $\score{q}{s_i}$ in order to use them in our rank aggregation function, together with proper normalization steps.
The notation $\scorein{\rankof{q}}{s_i}{s_j}$ stands for the general case, which is the score between $s_i$ and $s_j$ with respect to the same descriptor and comparator from the ranker that produced the rank $\rankof{q}$ for the query $q$.

While a ranker establishes a ranking model, different descriptors and comparators can be used to compose rankers, and it is well known that descriptors can be complementary, as well as comparators. Given a set of $m$ rankers, $\{R_1,R_2,\ldots,R_m\}$, being used for query retrieval over a collection $S$, for every query $q$ we can obtain $\ranksof{q}= \{\rankof{1},\rankof{2},\ldots,\rankof{m}\}$, from which a \textit{rank aggregation} function $f$ can produce a combined rank $\rankof{q,f}=f(\ranksof{q})$, hopefully more effective than the individual ranks $\rankof{1}$, $\rankof{2}$, etc.

\changed{We summarize the notations and their definitions in Table~\ref{tab:notations}.}

\begin{table}[ht]
\centering
\caption{Notations followed in our Retrieval Model.}
\label{tab:notations}
\resizebox{.99\columnwidth}{!}{\begin{tabular}{ll}
\hline
\textbf{Notation} & \textbf{Meaning} \\
\hline
$S$ & collection, or the response set in the context of search \\
$n$ & $|S|$ \\
$s$ & a sample, from $S$ \\
$\descriptor$ & descriptor \\
$\epsilon(s)$ & a data structure, generated by $\descriptor$, that describes $s$ \\
$\comparator$ & comparator \\
$\scoresymbol$ & score, of codomain $\mathbb{R^+}$, generated by $\comparator$ over $(\epsilon(s_i),\epsilon(s_j))$ \\
$q$ & query \\
$L$ & cut-off parameter \\
$\ranker$ & ranker, a tuple $(\descriptor,\comparator)$ \\
$\rank$ & rank, a permutation of $S_L \subseteq S$, generated by $\ranker$ \\
$\rankof{q}$ & rank for $q$ \\
$\scorein{\rankof{q}}{s_i}{s_j}$ & score between $s_i$ and $s_j$ with respect to the same $\ranker$ that produced $\rankof{q}$ for $q$ \\
$\rankposition{\rankof{q}}{x}$ & position of $x$ in $\rankof{q}$ \\
$m$ & number of rankers used \\
$\ranksof{q}$ & rank set for $q$, $\{\rankof{1},\rankof{2},\ldots,\rankof{m}\}$, generated by $\{R_1,R_2,\ldots,R_m\}$ \\
$f$ & a rank aggregation function \\
$\rankof{q,f}$ & the output rank of $f$, expressed by $f(\ranksof{q})$ \\
\hline
\end{tabular}}
\end{table}

\section{Unsupervised Graph-based Rank Aggregation Approach}
\label{propMethod}

We propose a graph-based rank aggregation function $f$ that works for any collection $S$ combined with the use of $m$ rankers of any kind. It relies on a {\em composite} ranker, whose descriptor extracts a graph-based representation, named {\em fusion graph}, from collection samples, and a fusion graph comparator is employed in this ranker. A fusion graph encodes contextual information from different ranks, defined in terms of multiple base rankers.

Both a query $q$ and each sample $s$ of a target collection are represented by graphs, say query fusion graph $G_{\ranksof{q}}$ and fusion graph $G_{\ranksof{s}}$. A search is, therefore, modeled as the ranking of graphs $G_{\ranksof{s}}$ of collection samples with respect to a query graph $G_{\ranksof{q}}$, i.e., the rank aggregation function $f$ is able to rank fusion graphs based on their similarity to a query graph.

\begin{figure}[!t]
\includegraphics[width=\linewidth]{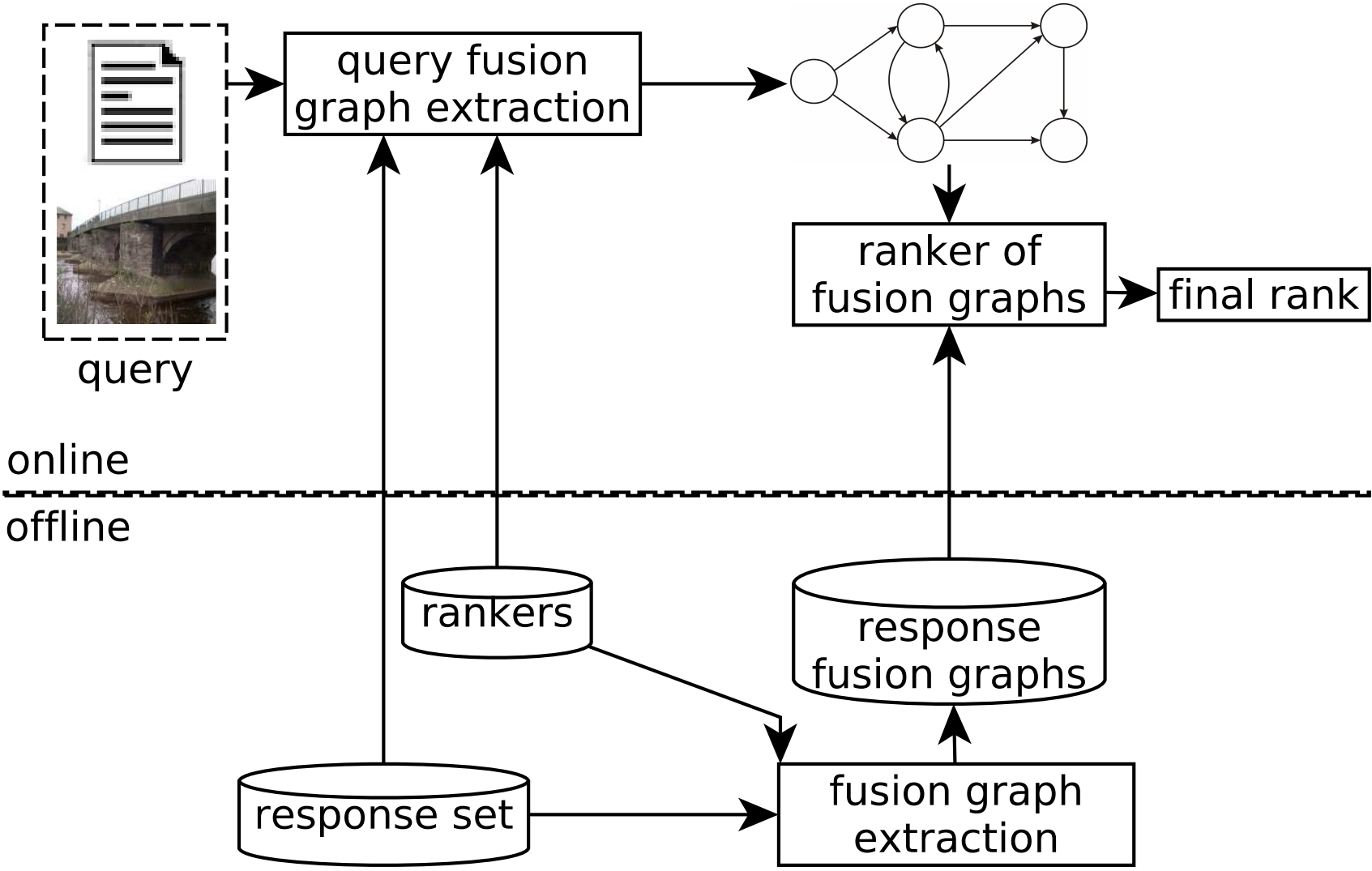}
\caption{Schematic view of the unsupervised graph-based rank aggregation approach.}\label{fig:graphFusionRetrieval}
\end{figure}

Figure~\ref{fig:graphFusionRetrieval} provides \changed{a} schematic overview of the unsupervised graph-based rank aggregation approach, which is composed of offline and online workflows. The steps `fusion graph extraction' and `ranker of fusion graphs' are detailed in Sections~\ref{sec:extractionFusionGraph} and~\ref{sec:retrFusionGraph}, respectively.
The offline workflow is responsible for representing the response set as fusion graphs, while the online workflow, in turn, processes a query and produces a final rank to be returned as the final result.

\subsection{Extraction of Fusion Graphs}
\label{sec:extractionFusionGraph}

\begin{figure}
\includegraphics[width=\linewidth]{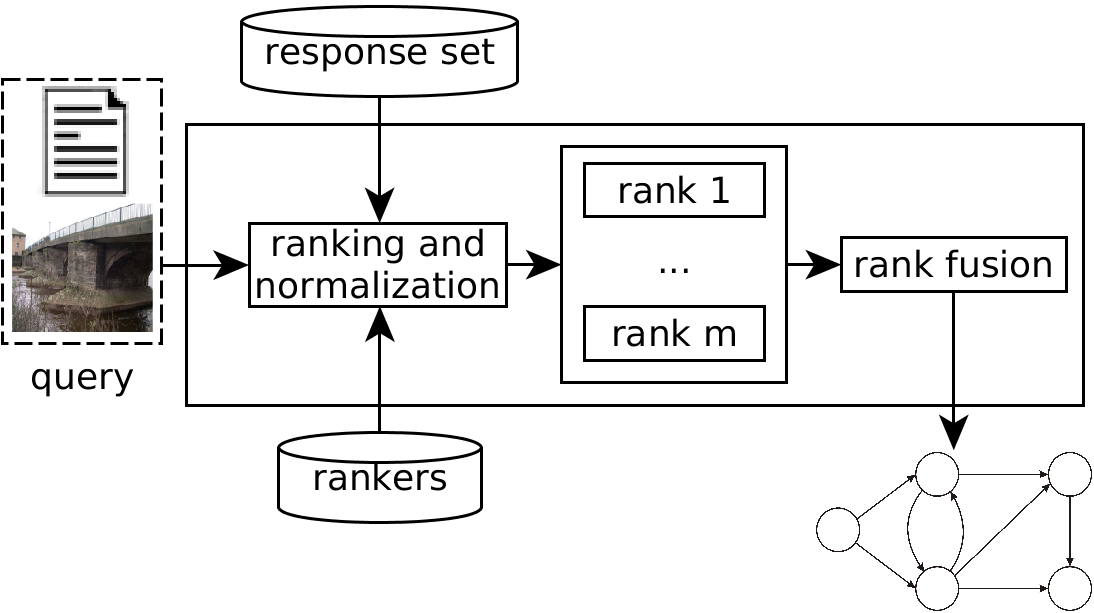}
\caption{Extraction of a fusion graph.}\label{fig:graphFusionFromSample}
\end{figure}

In both offline and online workflows, illustrated in Figure~\ref{fig:graphFusionRetrieval}, a fusion graph extraction step is adopted.
A fusion graph extraction aggregates ranks for a query, based on the rankers and response set used, producing a fusion graph per query.
It is basically comprised of three steps: creation of ranks using different rankers, rank normalization, and rank fusion. These components are illustrated in Figure~\ref{fig:graphFusionFromSample}.
The creation of ranks follows what was described in Section~\ref{secRetrModel}. Our rank aggregation function works upon a predefined set of rankers, so we assume that the base ranks for any query can be provided as requested.
Sections~\ref{sec:rankNormalization} and \ref{sec:rankFusion} detail the other steps.

\subsubsection{Rank Normalization}
\label{sec:rankNormalization}

For a certain ranker, its comparator~$\comparator$ may be either a distance or similarity function. Furthermore, different comparators may produce scores at different ranges.
Nevertheless, these scores are employed in our rank aggregation function.
For this reason, we need to normalize the comparator outputs, from the rankers being used, so that the scores from ranks become comparable. The ranks' scores must also fit into an uniform positive range, due to the way we use the scores in our rank aggregation formulation.

Assuming ranks of size $L$, we adopt a rank normalization procedure that relies on two steps: rank repositioning based on mutual and reciprocal rank references, and score \changed{rescaling}.

\changed{Rank relationships are not symmetric, in the sense that an object $i$ well ranked for a query $j$ does not imply that $j$ is well ranked for a query $i$.} However, improving the symmetry of the $k$-neighborhood usually improves the effectiveness of retrieval functions~\cite{Jegou:2010:contextualDissimilarity}.
\changed{In order to explore this behavior,}
we propose a rank repositioning, based on a neighborhood-aware distance $\delta$ given by Equation~\ref{eq:rankRepositioning}, where $\rankposition{\rankof{i}}{j} \le L$ and refers to the position of $j$ in the rank $\rankof{i}$. It considers both mutual~\cite{pedronette:2013:rlsim} and reciprocal~\cite{qin:2011:objRetrieval} neighborhood, and the ranks are then updated by a stable sorting algorithm over $\delta$, up to the top-$L$ positions.
\changed{The idea is to bring a ranked item $i$ to the top positions of the rank of $j$ as much as $j$ also has $i$ in top positions of its own rank.
The mutual neighborhood sums rank positions from both ranks, and the reciprocal neighborhood
considers only the maximum.}

\begin{equation}
\label{eq:rankRepositioning}
\delta(i,j) = \rankposition{\rankof{i}}{j} + \rankposition{\rankof{j}}{i} + max(\rankposition{\rankof{i}}{j}, \rankposition{\rankof{j}}{i})
\end{equation}	

For the second rank normalization step, we perform score \changed{rescaling} for the rank, assigning a uniform range from $1$, to the top-ranked response item, to $0.1$, to the top-$L$ ranked response item, adopting uniform steps within this interval.

\subsubsection{Rank Fusion}
\label{sec:rankFusion}

This step is responsible for producing graphs that reflect the ranks for query samples.
At first, in an offline stage, for each sample $s \in S$, we perform a search using $s$ as $q$ and obtain its corresponding set $\ranksof{q}$ of ranks, using a cut-off of $L$. 

The choice of $L$ depends on the intended result size. 
Due to the way we construct the fusion graph, especially the vertex and edge weights, the value of $L$ does not directly affect the quality of the model, not demanding empirical adjustment. The effect of changing the value of $L$ is to increase the effectiveness of the method, up to a certain limit. In practice, the choice of $L$ is guided only by the trade-off between efficiency and effectiveness.
Our method has only this parameter, as opposed to some related works~\cite{zhang:2015:queryRankFusion, pedronette:2017:reciprocalGraph}, usually dependent on multiple hyperparameters.

From $\ranks$, we derive a weighted directed graph $G_\ranks = (V_\ranks,E_\ranks)$ that combines information from the ranks of $\ranks$, where $V_\ranks$ is the vertex set and $E_\ranks$ is the edge set.
\changed{A fusion graph aims to be a discriminative and comparable representation of objects, based on their ranks and existing relations among ranks.
In this way, a fusion graph $G_{\ranksof{q}}$ of an object $q$ includes all response items from each rank $\rankof{q} \in \ranksof{q}$, as vertices. Vertices are connected by taking into account the degree of relationship between them, and the degree of their relationships to $q$.}

Algorithm~\ref{lst:rankFusion} illustrates how $G_\ranks$ is computed.
A vertex $v_A$ is associated with a collection sample $A$.
The vertex set is composed of the union of all samples found in all ranks defined for query $q$.
The weight of vertex $v_A$, $w_{v_A}$, is the sum of the similarity similarities that the response item $A$ has in the ranks of $q$ (lines $5$ to $10$, Equation~\ref{eq:Wv}).
\changed{The vertex weight is expected to encode how relevant a response item $A$ is to $q$}.

Edges are created to express the relationship between response items (lines $11$ to $20$).
There will be an edge $e_{A,B}$, linking $v_A$ to $v_B$, if $A$ and $B$ are both responses in any rank of $q$ and if $B$ occurs in any rank of $A$.
The weight of $e_{A,B}$, $w_{e_{A,B}}$, is the sum of the similarities that the response item $B$ has in the ranks of $A$, divided by the position of $A$ in each rank of $q$ (Equation~\ref{eq:We}), considering position values starting by $1$.
\changed{The scores of $B$ in each $\rankof{A}$ matters, so we sum them. Also, we weight these scores inversely to the position in which $A$ appears in $\rankof{q}$. The goal is to ensure that the weight of the edge between $A$ and $B$ also encodes the importance of $A$ with respect to $q$.}

\begin{lstlisting}[style=floatcode,language=Python,caption={Rank fusion.},label={lst:rankFusion},mathescape]
# inputs: ranks $\ranks{q}$, for the query $q$
# output: a weighted directed graph $G_\ranks$
$G_\ranks$ = WeightedDirectedGraph() # $(V_\ranks,E_\ranks)$
for $\rank$ in $\ranksof{q}$:  # create vertices
	for A in $\rank$:
		weight $= \scorein{\rank}{q}{A}$
		if $v_A \not\in V_\ranks$:  # if new vertex
			$V_\ranks = V_\ranks \cup v_A$
			$w_{v_A} =$ weight
		else: $w_{v_A} \mathrel{+}=$ weight
for $\rank$ in $\ranksof{q}$:  # create edges
	for A in $\rank$:
		for $\rankof{A}$ in $\ranksof{A}$:
			for B in $\rankof{A}$:
				if $v_B \in V_\ranks$ and A != B:
					weight $= \scorein{\rankof{A}}{A}{B} \div \rankposition{\rank}{A}$
					if $e_{A,B} \not\in E_\ranks$:  # if new edge
						$E_\ranks = E_\ranks \cup e_{A,B}$
						$w_{e_{A,B}} =$ weight
					else: $w_{e_{A,B}} \mathrel{+}=$ weight
g.normalizeWeights(0,1)
\end{lstlisting}

\begin{figure}[!t]
\includegraphics[width=\linewidth]{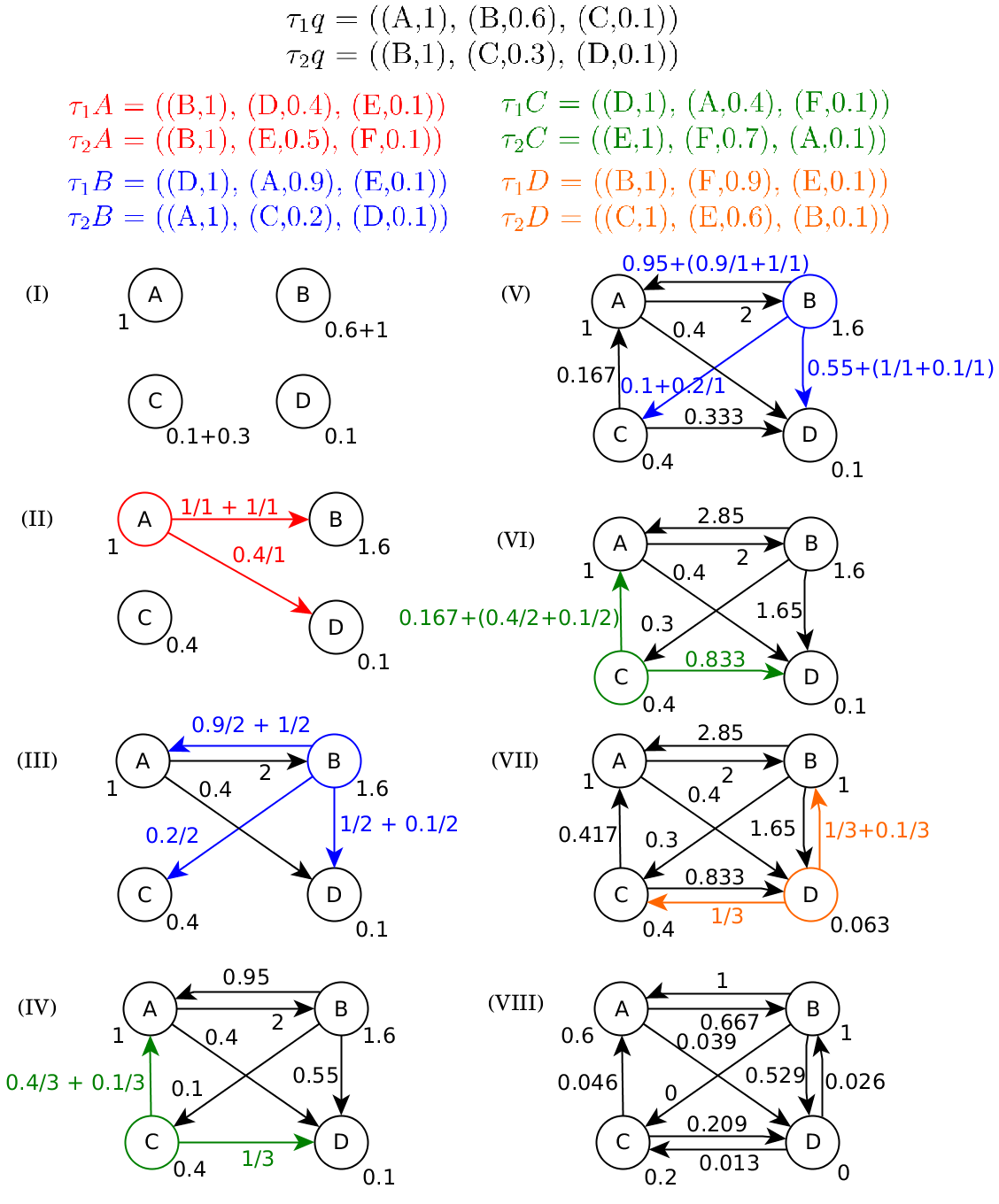}
\caption{\changed{Example of graph construction during rank fusion. The scores are shown along with the response items, within the ranks, for simplicity.}}\label{fig:sampleFusion}
\end{figure}

\begin{equation}
\label{eq:Wv}
	w_{v_A} = \sum_{A \in \rankof{i} \wedge \rankof{i} \in \ranksof{q}} \scorein{\rankof{i}}{q}{A}
\end{equation}	

\begin{equation}
\label{eq:We}
	w_{e_{A,B}} = \sum_{ A \in \rankof{i} \wedge \rankof{i} \in \ranksof{q}} \sum_{B \in \rankof{j} \wedge \rankof{j} \in \ranksof{A}} \left( \scorein{\rankof{j}}{A}{B} \div \rankposition{\rankof{i}}{A} \right)
\end{equation}

The creation of $G_\ranks$ ends with a weight normalization (line $21$), which makes the graph comparable by means of a graph comparator.
The weight of each vertex $v_i$, $w_{v_i}$, is replaced by $\frac{w_{v_i}}{max(w_v)}$, and the weight of edge $e_{i,j}$, $w_{e_{i,j}}$, is replaced by $\frac{w_{e_{i,j}}}{max(w_e)}$.

Figure~\ref{fig:sampleFusion} illustrates an example for the rank fusion, assuming a query $q$ and the use of two rankers. At first, a fully disconnected graph is built based on the retrieved results and their scores (I). Then, the relationships between the results (encoded in the ranks for B in blue, C in green, and D in orange), in their own ranks, are propagated into the graph as edges (see III, IV, and V). The resulting graphs, VI and VII in Figure~\ref{fig:sampleFusion}, correspond to the final fusion graph before and after normalization, respectively.

\subsection{Retrieval based on Fusion Graphs}
\label{sec:retrFusionGraph}

Our proposed rank aggregation function $f$ relies on a {\em composite} ranker that is defined as $\ranker_G = (D_G,\comparator_G)$, where $D_G$ is a descriptor which extracts fusion graphs, and $\comparator_G$ is a fusion graph comparator.
Given two fusion graphs $G_a$ and $G_b$, $\score{G_a}{G_b}$ can be computed by $\comparator_G$ using any graph-based similarity or dissimilarity function.
We propose the adoption of formulations based on the minimum common subgraph (\textit{mcs}), such as MCS~\cite{Bunke:1998:MCS} or WGU~\cite{Wallis:2001:WGU}. 
A graph $M$ is the \textit{mcs} of two weighted graphs $G_a$ and $G_b$ if:
(1) $M \subseteq G_a$
(2) $M \subseteq G_b$
and (3) there is no other subgraph $M'$ ($M' \subseteq G_a$, $M' \subseteq G_b$), such that $|M'| > |M|$, where $|M|$ is given by the sum of the vertex weights and edge weights of $M$.

MCS and WGU are shown in Equation~\ref{eq:MCS} and~\ref{eq:WGU}, respectively. In MCS, the larger the $|mcs|$ is, the more similar the two graphs are, which decreases the distance up to $0$. This metric produces values in [0, 1].
WGU behaves similarly to MCS with respect to identical graphs or graphs without intersection, and also produces values in [0, 1].
The denominator in WGU represents the size of the union of the two graphs, whose motivation is to allow for changes in the smaller graph to influence the distance value, which is not covered in MCS.
\begin{equation} \label{eq:MCS}
dist_{MCS(G_a,G_b)} = 1 - \frac{|mcs(G_a,G_b)|}{max(|G_a|,|G_b|)}
\end{equation}
\begin{equation} \label{eq:WGU}
dist_{WGU(G_a,G_b)} = 1 - \frac{|mcs(G_a,G_b)|}{|G_a| + |G_b| - |mcs(G_a,G_b)|}
\end{equation}

Note that the scores $\score{G_{\ranksof{q}}}{G_{\ranksof{s_i}}}$ and $\score{G_{\ranksof{q}}}{G_{\ranksof{s_j}}}$ can be compared to infer whether $s_i$ or $s_j$ is more relevant to $q$. The higher the score, the most similar the query and the response item are, regarding their ranks. A fusion graph, therefore, is able to encode intrinsic contextual information from multiple ranks.

The rank aggregation function $f$ is defined as $f(\ranksof{q}) = \rankof{q,f} = \{s_1,s_2,\ldots,s_n\}$ such that $|\rankof{q,f}| \leq L$ and $\{\score{G_{\ranksof{q}}}{G_{\ranksof{s_1}}},\score{G_{\ranksof{q}}}{G_{\ranksof{s_2}}}, \ldots, \score{G_{\ranksof{q}}}{G_{\ranksof{s_n}}}\}$ is in increasing order.

\subsection{Computational Cost Analysis}
\label{computationalAnalysis}

The cost of executing a query $q$ is the sum of the costs for: generating the ranks $\ranksof{q}$ for $q$; generating the query fusion graph $G_{\ranksof{q}}$; and retrieving samples using $G_{\ranksof{q}}$ and response set fusion graphs.

For the first part, individual ranks $\rank \in \ranksof{q}$ can be generated in parallel. Therefore, this step is limited by the cost of the slowest ranker.
Note that, typically, rankers adopt indexing structures, such as KD-tree or inverted files, leading to rank generations in sub-linear time with respect to the response set.

For the second part, the Rank fusion (Algorithm~\ref{lst:rankFusion}) has an asymptotic cost of $O(mL)$ for the first outer loop (create vertices), and $O(mLmL)$ for the second outer loop (create edges), leading to a total cost of $O(m^2L^2)$. As we use small values of $L$ and also a small number ($m$) of rankers, the cost of the rank fusion algorithm itself is negligible when compared to the step concerned with the generation of ranks.

In our fusion graph, vertices have different labels, and therefore we use graph comparison functions that take advantage of efficient algorithms for computing minimum common subgraphs, reducing its asymptotic cost to $O(|V_1||V_2|)$~\cite{dickinson2004matching}. Note also that the number of vertices in each graph is $O(mL)$ in the worst case. Both aspects lead to efficient graph comparators, in practice.
The graph-based retrieval can be either implemented linearly over the response set, or sub-linearly using indexing methods such as graph embedding techniques~\cite{Dourado:2019:BoTG,Silva:2018:BoG,Zhu:2014}. We leave this investigation for future work.

The graph-based retrieval, for each $q$, relies on the existence of ranks and fusion graphs from the response set, but both steps are performed only once per collection (`fusion graph extraction' in Figure~\ref{fig:graphFusionRetrieval}), and in an offline stage. 

\section{Experimental Evaluation}
\label{secExpEval}

This section presents the adopted evaluation protocol and experimental results related to the comparison of our method with individual rankers and other rank aggregation approaches.

\subsection{Datasets and features}

We selected datasets of different purposes, compositions, and sizes in order to validate our method in different searching scenarios. Table~\ref{tab:datasets} lists the datasets used, and Table~\ref{tab:rankers} summarizes the individual rankers being adopted per dataset in order to (1) be evaluated in isolation, and (2) generate rankers for the fusions.

\begin{table}[!t]
\centering
\caption{Datasets used in the experimental evaluation.}
\label{tab:datasets}
\begin{tabular}{lrl}
\hline
\textbf{Dataset} & \textbf{Size} & \textbf{Type} \\
\hline
Ohsumed & \num{34389} & Textual \\
Brodatz & \num{1776} & Texture \\
MPEG-7 & \num{1400} & Shape \\
Soccer & \num{280} & Color Scenes \\
UW & \num{1109} & Color Scenes and Keywords \\
UKBench & \num{10200} & Objects / Scenes \\
\hline
\end{tabular}
\end{table}

\begin{table*}[!t]
\centering
\caption{Individual rankers adopted per dataset in the experimental evaluation.}
\label{tab:rankers}
\begin{tabular}{lp{6.9cm}p{2.4cm}}
\hline
\textbf{Dataset} & \textbf{Rankers} & \textbf{Type} \\ \hline
Ohsumed & BoW-cosine, BoW-Jaccard, 2grams-cosine, 2grams-Jaccard, GNF-MCS, GNF-WGU, WMD & Textual \\ \hline
Brodatz & LBP, CCOM, LAS & Texture \\\hline
MPEG-7 & SS, BAS, IDSC, CFD, ASC, AIR & Shape \\ \hline
Soccer & GCH, ACC, BIC & Color \\\hline
UW & GCH, BIC, JAC, HTD, QCCH, LAS, COSINE, JACCARD, TF-IDF, DICE, OKAPI, BOW & Color, Texture, Textual \\\hline
UKBench & ACC, VOC, SCD, JCD, CNN-Caffe, FCTH-SPy, CEDD-SPy & Color, Texture, BoVW, CNN \\\hline
\end{tabular}
\end{table*}

\textbf{Ohsumed}~\cite{hersh:1994:ohsumed} is a textual dataset, composed of bibliographic medical documents, provided by the National Library of Medicine. It contains \num{34389} cardiovascular diseases abstracts, distributed across \num{23} Medical Subject Headings (MeSH) diseases categories of cardiovascular diseases group. Without loss of generality, we used the subset of \num{18302} uni-labeled documents, varying from 56 to 2876 documents per category.
For Ohsumed, we adopted $7$ rankers\footnote{For all textual rankers used in the experimental evaluation, we preprocess the documents with stop word removal and Porter's stemming.}:
\begin{itemize}
\item $2$ using the Bag-of-Words (BoW)%
, with comparators cosine and Jaccard: BoW-cosine and BoW-Jaccard;
\item $2$ using the 2grams descriptor, with comparators cosine and Jaccard; 2grams-cosine and 2grams-Jaccard;
\item $2$ using a graph-based descriptor, called \textit{normalized-frequency} (GNF)~\cite{schenker2007:graphtext}, with comparators MCS~\cite{Bunke:1998:MCS} and WGU~\cite{Wallis:2001:WGU}: GNF-MCS and GNF-WGU;
\item WMD~\cite{kusner:2015:WMD}, a ranker based on \textit{word embeddings}~\cite{mikolov:2013:wordEmbedding}.
\end{itemize}

\textbf{Brodatz}~\cite{brodatz:1966} is a texture dataset. There are \num{1776} images (texture blocks), being $16$ samples for each of the $111$ classes (texture types). We adopt $3$ texture rankers: Local Binary Patterns (LBP)~\cite{ojala:2002:LBP},
Color Co-Occurrence Matrix
(CCOM)~\cite{kovalev:1998:CCOM},
and Local Activity Spectrum~(LAS)~\cite{tao:2000:LAS}.

\textbf{MPEG-7}~\cite{latecki:2000:shape} is a shape dataset, composed of $1400$ images, equally distributed in $20$ images per $70$ categories. We adopt $6$ shape rankers:
Segment Saliences (SS)~\cite{torres:2007:SS},
Beam Angle Statistics (BAS)~\cite{arica:2003:BAS},
Inner Distance Shape Context (IDSC)~\cite{ling:2007:IDSC},
Contour Features Descriptor (CFD)~\cite{pedronette:2010:CFD},
Aspect Shape Context (ASC)~\cite{ling:2010:ASC}, and
Articulation-Invariant Representation (AIR)~\cite{gopalan:2010:AIR}.

\textbf{Soccer}~\cite{van:2006:soccer} is an image dataset, composed of $280$ images, equally distributed in $40$ images per $7$ categories (the soccer teams).
We adopt 3 color-based rankers:
Global Color Histogram (GCH)~\cite{swain:1991:GCH},
Auto Color Correlograms (ACC)~\cite{huang:1997:ACC}, and
Border/Interior Pixel Classification (BIC)~\cite{stehling:2002:BIC}.

University of Washington (\textbf{UW})~\cite{Deselaers:2008:UW} is a hybrid dataset, composed of $1109$ pictures from different locations, annotated by textual keywords. The number of keywords per picture vary from $1$ to $22$. There are $20$ classes, varying from $22$ to $255$ pictures per class.
We adopt $12$ rankers, comprising $3$ types:
\begin{itemize}
\item 3 Visual color rankers: GCH~\cite{swain:1991:GCH}, BIC~\cite{stehling:2002:BIC}, and Joint Autocorrelogram (JAC)~\cite{williams:2007:JAC};
\item 3 Visual texture rankers: Homogeneous Texture Descriptor (HTD)~\cite{wu:1999:HTD}, Quantized Compound Change Histogram (QCCH)~\cite{huang:2007:QCCH}, and Local Activity Spectrum (LAS)~\cite{tao:2000:LAS};
\item 6 Textual rankers:
COSINE~\cite{baeza:1999:modernIR},
JACCARD~\cite{lewis2006textsimilarityan},
Term Frequency-Inverse Document Frequency (TF-IDF)~\cite{baeza:1999:modernIR},
DICE~\cite{lewis2006textsimilarityan},
OKAPI~\cite{robertson:1995:okapi}, and
BOW~\cite{carrillo2009representing}.
\end{itemize}

\textbf{UKBench}~\cite{nister:2006:UKBench} is a dataset of \num{10200} images, consisting of \num{2550} scenes/objects captured $4$ times each. The captures vary in terms of viewpoint, illumination, and distance. The objects/scenes correspond to the categories, so there are four samples per class. Due to the small and fixed category sizes, effectiveness assessment using this dataset  relies on an evaluation metric, called N-S Score, varying from $1$ to $4$, which measures the mean number of relevant images among the first four images retrieved.
We adopt seven rankers, based on color  and texture properties. Some of them are based on global descriptors, while others rely on local features:
\begin{itemize}
\item ACC~\cite{huang:1997:ACC};
\item Vocabulary Tree (VOC)~\cite{wang:2011:VOC}, that uses SIFT;
\item CNN-Caffe~\cite{Jia:2014:CNNCaffe}: features extracted from the 7th layer of a Convolution Neural Network (CNN) obtained with the Caffe framework. A 4096-dimensional descriptor is extracted per image, and the Euclidean distance is used as the comparator.
\item Scalable Color Descriptor (SCD)~\cite{manjunath:2001:SCD}
\item Joint Composite Descriptor (JCD)~\cite{zagoris:2010:JCD}
\item Fuzzy Color and Texture Histogram Spatial Pyramid (FCTH-SPy)~\cite{chatzichristofis:2008:FCTH,lux:2011:irLIRe}
\item Color and Edge Directivity Descriptor Spatial Pyramid (CEDD-SPy)~\cite{chatzichristofis:2008:CEDD,lux:2011:irLIRe}
\end{itemize}

\begin{table*}
\caption{Results for individual rankers on textual, image, and hybrid datasets.}
\begin{subtable}[t]{0.32\textwidth}
\centering
\caption{Brodatz}
\label{tab:rankersResultsBrodatz}
\begin{tabular}{lr}\hline
Ranker & NDCG@10  \\\hline
LAS        & 0.850533 \\
CCOM       & 0.726186 \\
LBP        & 0.652759 \\\hline
\end{tabular}
\vspace{2mm}
\caption{UW dataset}
\label{tab:rankersResultsUW}
\resizebox{3.9cm}{!}{%
\begin{tabular}{lr}\hline
Ranker & NDCG@10  \\\hline
JAC & 0.810729 \\
BIC & 0.746454 \\
DICE & 0.722831 \\
BOW & 0.720781 \\
OKAPI & 0.716035 \\
JACCARD & 0.701651 \\
TF-IDF & 0.658880\\
GCH & 0.630315 \\
COSINE & 0.554767 \\
LAS & 0.514314 \\
HTD & 0.495002 \\
QCCH & 0.414249 \\\hline
\end{tabular}}
\end{subtable}
\begin{subtable}[t]{0.32\textwidth}
\centering
\caption{MPEG-7}
\label{tab:rankersResultsMPEG7}
\resizebox{3.9cm}{!}{%
\begin{tabular}{lr}\hline
Ranker & NDCG@10  \\\hline
ASC  & 0.941585 \\
AIR  & 0.939424 \\
CFD  & 0.930685 \\
IDSC & 0.922828 \\
BAS  & 0.866098 \\
SS   & 0.611481 \\\hline
\end{tabular}}
\vspace{2mm}
\caption{Ohsumed}
\label{tab:rankersResultsOhsumed}
\resizebox{3.9cm}{!}{%
\begin{tabular}{lr}\hline
Ranker & NDCG@10  \\\hline
BoW-cosine & 0.669701 \\
2grams-cosine & 0.664120 \\
GNF-WGU & 0.662668 \\
GNF-MCS & 0.655420 \\
2grams-Jaccard & 0.651320 \\
BoW-Jaccard & 0.645711 \\
WMD & 0.427361 \\\hline
\end{tabular}}
\end{subtable}
\begin{subtable}[t]{0.32\textwidth}
\centering
\caption{UKBench}
\label{tab:rankersResultsUKBench}
\begin{tabular}{lr}\hline
Ranker & N-S Score \\\hline
VOC & 3.54 \\
ACC & 3.37 \\
CNN-Caffe & 3.31 \\
SCD & 3.15 \\
JCD & 2.79 \\
FCTH-SPy & 2.73 \\
CEDD-SPy & 2.61 \\
\hline
\end{tabular}
\vspace{2mm}
\caption{Soccer}
\label{tab:rankersResultsSoccer}
\begin{tabular}{lr}\hline
Ranker & NDCG@10  \\\hline
BIC & 0.614818 \\
ACC & 0.592699 \\
GCH & 0.536412 \\\hline
\end{tabular}
\end{subtable}
\end{table*}

\begin{table}
\centering
\caption{Correlation of individual ranks on Brodatz.}
\label{tab:correlationsBrodatz}
\begin{tabular}{llll}\hline
 & CCOM & LAS & LBP \\\hline
CCOM & 1.00 & 0.38 & 0.25 \\
LAS  & 0.38 & 1.00 & 0.30 \\
LBP  & 0.25 & 0.30 & 1.00 \\\hline
\end{tabular}
\end{table}

\begin{table*}
\centering
\caption{Correlation of individual ranks on UW.}
\label{tab:correlationsUW}
\resizebox{.99\columnwidth}{!}{\begin{tabular}{lllllllllllll}\hline
& BIC & GCH & HTD & JAC & LAS & QCCH & BOW & COSINE & DICE & JACCARD & OKAPI & TF-IDF \\\hline
BIC & 1.00 & 0.29 & 0.12 & 0.27 & 0.12 & 0.11 & 0.14 & 0.11 & 0.14 & 0.13 & 0.13 & 0.11 \\
GCH & 0.29 & 1.00 & 0.11 & 0.18 & 0.11 & 0.10 & 0.11 & 0.08 & 0.11 & 0.10 & 0.09 & 0.08 \\
HTD & 0.12 & 0.11 & 1.00 & 0.12 & 0.12 & 0.12 & 0.09 & 0.07 & 0.09 & 0.08 & 0.07 & 0.07 \\
JAC & 0.27 & 0.18 & 0.12 & 1.00 & 0.11 & 0.10 & 0.14 & 0.11 & 0.15 & 0.14 & 0.13 & 0.12 \\
LAS & 0.12 & 0.11 & 0.12 & 0.11 & 1.00 & 0.16 & 0.08 & 0.06 & 0.08 & 0.08 & 0.07 & 0.07 \\
QCCH & 0.11 & 0.10 & 0.12 & 0.10 & 0.16 & 1.00 & 0.08 & 0.06 & 0.08 & 0.07 & 0.06 & 0.06 \\
BOW & 0.14 & 0.11 & 0.09 & 0.14 & 0.08 & 0.08 & 1.00 & 0.22 & 0.44 & 0.42 & 0.30 & 0.25 \\
COSINE & 0.11 & 0.08 & 0.07 & 0.11 & 0.06 & 0.06 & 0.22 & 1.00 & 0.32 & 0.32 & 0.36 & 0.45 \\
DICE & 0.14 & 0.11 & 0.09 & 0.15 & 0.08 & 0.08 & 0.44 & 0.32 & 1.00 & 0.85 & 0.37 & 0.37 \\
JACCARD & 0.13 & 0.10 & 0.08 & 0.14 & 0.08 & 0.07 & 0.42 & 0.32 & 0.85 & 1.00 & 0.38 & 0.37 \\
OKAPI & 0.13 & 0.09 & 0.07 & 0.13 & 0.07 & 0.06 & 0.30 & 0.36 & 0.37 & 0.38 & 1.00 & 0.60 \\
TF-IDF & 0.11 & 0.08 & 0.07 & 0.12 & 0.07 & 0.06 & 0.25 & 0.45 & 0.37 & 0.37 & 0.60 & 1.00 \\\hline
\end{tabular}}
\end{table*}

\begin{table}
\centering
\caption{Correlation of individual ranks on MPEG-7.}
\label{tab:correlationsMPEG7}
\begin{tabular}{lllllll}\hline
 & AIR & ASC & BAS & CFD & IDSC & SS \\\hline
AIR & 1.00 & 0.31 & 0.27 & 0.30 & 0.30 & 0.18 \\
ASC & 0.31 & 1.00 & 0.33 & 0.37 & 0.70 & 0.20 \\
BAS & 0.27 & 0.33 & 1.00 & 0.48 & 0.32 & 0.28 \\
CFD & 0.30 & 0.37 & 0.48 & 1.00 & 0.36 & 0.26 \\
IDSC & 0.30 & 0.70 & 0.32 & 0.36 & 1.00 & 0.19 \\
SS & 0.18 & 0.20 & 0.28 & 0.26 & 0.19 & 1.00 \\\hline
\end{tabular}
\end{table}

\begin{table*}
\centering
\caption{Correlation of individual ranks on Ohsumed.}
\label{tab:correlationsOhsumed}
\resizebox{.99\columnwidth}{!}{\begin{tabular}{llllllll}\hline
& BoW-cosine & BoW-Jaccard & 2grams-cosine & 2grams-Jaccard & GNF-MCS & GNF-WGU & WMD \\\hline
BoW-cosine & 1.00 & 0.56 & 0.48 & 0.45 & 0.49 & 0.55 & 0.10 \\
BoW-Jaccard & 0.56 & 1.00 & 0.41 & 0.50 & 0.51 & 0.54 & 0.11 \\
2grams-cosine & 0.48 & 0.41 & 1.00 & 0.64 & 0.51 & 0.58 & 0.10 \\
2grams-Jaccard & 0.45 & 0.50 & 0.64 & 1.00 & 0.55 & 0.61 & 0.10 \\
GNF-MCS & 0.49 & 0.51 & 0.51 & 0.55 & 1.00 & 0.73 & 0.10 \\
GNF-WGU & 0.55 & 0.54 & 0.58 & 0.61 & 0.73 & 1.00 & 0.10 \\
WMD & 0.10 & 0.11 & 0.10 & 0.10 & 0.10 & 0.10 & 1.00 \\\hline
\end{tabular}}
\end{table*}

\begin{table}
\centering
\caption{Correlation of individual ranks on UKBench.}
\label{tab:correlationsUKBench}
\resizebox{.99\columnwidth}{!}{\begin{tabular}{llllllll}\hline
 & ACC & VOC & CNN-Caffe & SCD & JCD & FCTH-SPy & CEDD-SPy \\\hline
ACC & 1.00 & 0.23 & 0.22 & 0.31 & 0.23 & 0.22 & 0.21 \\
VOC & 0.23 & 1.00 & 0.24 & 0.22 & 0.21 & 0.20 & 0.20 \\
CNN-Caffe & 0.22 & 0.24 & 1.00 & 0.21 & 0.20 & 0.19 & 0.19 \\
SCD & 0.31 & 0.22 & 0.21 & 1.00 & 0.26 & 0.26 & 0.23 \\
JCD & 0.23 & 0.21 & 0.20 & 0.26 & 1.00 & 0.39 & 0.53 \\
FCTH-SPy & 0.22 & 0.20 & 0.19 & 0.26 & 0.39 & 1.00 & 0.28 \\
CEDD-SPy & 0.21 & 0.20 & 0.19 & 0.23 & 0.53 & 0.28 & 1.00 \\\hline
\end{tabular}}
\end{table}

\begin{table}
\centering
\caption{Correlation of individual ranks on Soccer.}
\label{tab:correlationsSoccer}
\begin{tabular}{llll}\hline
& BIC & ACC & GCH \\\hline
BIC & 1.00 & 0.46 & 0.27 \\
ACC & 0.46 & 1.00 & 0.30 \\
GCH & 0.27 & 0.30 & 1.00 \\\hline
\end{tabular}
\end{table}

\subsection{Experimental Procedure}

We evaluate our method, as well as the individual rankers and baselines, with respect to the effectiveness in retrieval tasks. \changed{For the Ohsumed dataset, we implemented the rankers and extracted the ranks ourselves. For the other datasets, we adopted ranks built from previous works of our research group~\cite{pedronette:2017:reciprocalGraph,pedronette:2016:CorGraph}.}

Due to the nature of the datasets used, we use each sample $s$ as query $q$ at a time, whose result candidates belong to $S$, and we consider a retrieved item as relevant to the query if it belongs to the same class of the query sample, since we are validating in labeled collections, i.e., relevant labels in the experiments are either $1$ for relevant or $0$ for irrelevant. Therefore, in this case, the query set size corresponds to the dataset size.
Separate query and response sets can be used, as well as graded relevance, but these aspects do not affect the applicability of our model. This protocol concerns document retrieval, also referred to as ad hoc retrieval, which was also very usual in validation protocol of our baselines.

We use normalized discounted cumulative gain at cutoff 10 (NDCG@10) for all datasets except for UKBench, for which we use the adopted N-S Score effectiveness measure, the standard measure used in this dataset.


For each dataset, we evaluate the effectiveness of the individual rankers, and also their correlation.
Both the effectiveness and the correlation scores are used to guide the choice of base rankers.
We evaluate three approaches for selecting rankers: all rankers available for each dataset;
the pair composed of the two best rankers in terms of effectiveness; and the pair of rankers that present the best balance between high effectiveness and low correlation.

The second and third approaches may lead to the use of the same pair of rankers. Therefore, in cases where this happens, we also present the aggregation using the three most effective rankers. For the third approach, we select the pair of rankers $R_x$ and $R_y$ that maximizes the selection measure $M(R_x,R_y)$ expressed in Equation~\ref{eq:rankerPairSelection}:
\begin{equation} \label{eq:rankerPairSelection}
M(R_x,R_y) = \frac{ 1 + ef_{R_x} \times ef_{R_y} }{ 1 + cor(R_x, R_y) }
\end{equation}

\noindent where $ef_{R_x}$ denotes the effectiveness value for the ranker $R_x$, regardless the evaluation metric used (NDCG@10 or N-S Score), and $cor(R_x,R_y)$ is the correlation between $R_x$ and $R_y$. This is a modified measure adapted from the one proposed in~\cite{valem2017selection}.


\changed{Let $\rankof{A}$ and $\rankof{B}$ be two ranks, and $n$ be the size of these ranks.}
The correlation between two rankers is given by the mean correlation of their ranks with respect to each query.
We adopt Jaccard's correlation, given by Equation~\ref{eq:JaccardCorrelation}.
Other metrics were considered,
\changed{but Jaccard was the one that achieved ranker combinations for rank aggregation with the best results, in preliminary analysis that we performed, considering the possibilities for computing $cor(R_x, R_y)$ from Equation~\ref{eq:rankerPairSelection} as with Jaccard, Kendall Tau or Spearman.
An equivalent conclusion was observed in~\cite{valem2017selection}, that investigated possibilities for ranker selection.}

\changed{Kendall Tau relies on the number of discordant pairs between $\rankof{A}$ and $\rankof{B}$.
Given two response items $(s_i,s_j)$, this pair is named discordant for $\rankof{A}$ and $\rankof{B}$, if $\rankposition{\ranksof{A}}{s_i} > \rankposition{\ranksof{B}}{s_j}$ and $\rankposition{\ranksof{A}}{s_j} > \rankposition{\ranksof{B}}{s_i}$.
Kendall Tau's correlation is given by Equation~\ref{eq:KendallTauCorrelation}, where $K_d$ is the number of discordant pairs and $n_d = \frac{ n \times (n-1) }{ 2 }$.
Spearman correlation relies on the position disparity of each response item in the two ranks, and it is given by Equation~\ref{eq:SpearmanCorrelation}.}

\begin{equation} \label{eq:JaccardCorrelation}
J(\rankof{A}, \rankof{B}) = \frac{ \mid \rankof{A} \cap \rankof{B} \mid }{ \mid \rankof{A} \cup \rankof{B} \mid }
\end{equation}
\changed{%
\begin{equation} \label{eq:KendallTauCorrelation}
K_s(\rankof{A}, \rankof{B}) = 1 - \frac{ K_d( \rankof{A}, \rankof{B}) }{ n_d }
\end{equation}
}
\changed{%
\begin{equation} \label{eq:SpearmanCorrelation}
S(\rankof{A}, \rankof{B}) = 1 - \frac{ \sum_{s_i \in \ranksof{A}} \mid \rankposition{\ranksof{A}}{s_i} - \rankposition{\ranksof{B}}{s_i} \mid }{ n \times (n+1) }
\end{equation}
}

\changed{Several state-of-the-art rank aggregation baselines are tested, along with our method, for the same candidate set of rankers:
QueryRankFusion~\cite{zhang:2015:queryRankFusion},
RecKNNGraphCCs~\cite{pedronette:2017:reciprocalGraph},
RkGraph~\cite{pedronette:2016:RkGraph},
CorGraph~\cite{pedronette:2016:CorGraph},
MRA~\cite{fagin:2003:MRA},
RRF~\cite{cormack2:2009:RRF},
CombSUM~\cite{fox:1994:combination}, CombMIN, CombMAX, CombMED, CombANZ, CombMNZ,
BordaCount~\cite{borda:1784:bordaCount},
Condorcet,
Kemeny,
and RLSim~\cite{pedronette:2013:rlsim}.
These baselines were detailed in Section~\ref{secRW}.
They are unsupervised rank aggregation methods, as it is our method, and they cover most state-of-the-art graph-based approaches, as well as some classic but still competitive ones.
Because we propose an unsupervised method, we adopted unsupervised baselines to make fair comparisons.}
For UKBench, we also compare the results with the ones associated with the methods described in the following recent works:
\citet{bai:2016:sparse},
\citet{xie:2015:image},
\citet{zheng:2015:query},
\citet{zheng:2014:lpnorm},
\citet{wang:2012:metricfusion},
and \citet{qin:2011:objRetrieval}.

We conduct statistical tests, using per-query paired t-test at 99\% confidence level.
We denote the statistical analysis with the following symbols:
$\blacktriangle$ indicates that our method was statistically better than the baseline, $\blacktriangledown$ means the opposite, and $\bullet$ means a statistical tie.

As we analyze a large number of datasets, fusion configurations (which rankers to fuse) and baselines, besides the statistical comparisons we also present the \textit{winning number}~\cite{Tax:2018:benchmarkLtK}
of each rank aggregation function, aiming at providing a global performance indicator per method.
The winning number of a method $m$, $W_m$, regarding a performance measure $P$, is adapted to our context as in Equation~\ref{eq:winningNumber}, where
$D$ is the set of datasets,
$C_d$ is the set of our $3$ pre-defined configurations for dataset $d$ with respect to which rankers to fuse,
$P_m(d,c)$ is the performance of the method $m$ on dataset $d$ and configuration $c \in C_d$,
$M$ is set of rank aggregation methods, and $\bm{1}_{P_m(d) > P_k(d)}$ is the indicator function given by Equation~\ref{eq:indicatorFunction}.

\begin{equation} \label{eq:winningNumber}
W_m = \sum_{d \in D} \sum_{c \in C_d} \sum_{i \in M} 1_{P_m(d,c) > P_i(d,c)}
\end{equation}

\begin{equation}\label{eq:indicatorFunction}
\bm{1}_{P_m(d,c) > P_i(d,c)} =
    \begin{cases} 
      1 & \text{if $P_m(d,c) > P_i(d,c)$}, \\
      0 & \text{otherwise.}
\end{cases}
\end{equation}

\subsection{Ranker Effectiveness and Correlations}

Tables~\ref{tab:rankersResultsBrodatz},
\ref{tab:rankersResultsUW},
\ref{tab:rankersResultsMPEG7},
\ref{tab:rankersResultsOhsumed},
\ref{tab:rankersResultsUKBench}, and
\ref{tab:rankersResultsSoccer} report the results obtained by the individual rankers, respectively for the datasets Brodatz, UW, MPEG-7, Ohsumed, UKBench, and Soccer.
The rankers are presented sorted by their results.
It can be noticed large variability in rankers' results.
Furthermore, rankers perform differently depending on the dataset, possibly providing complementary views. For example,
JACCARD was better than COSINE in UW, but the opposite happened for the Ohsumed dataset.

Tables~\ref{tab:correlationsBrodatz},
\ref{tab:correlationsUW},
\ref{tab:correlationsMPEG7},
\ref{tab:correlationsOhsumed},
\ref{tab:correlationsUKBench}, and
\ref{tab:correlationsSoccer} report the Jaccard's correlations between ranks for the individual rankers used, respectively for the datasets Brodatz, UW, MPEG-7, Ohsumed, UKBench, and Soccer.
These correlations, along with the individual rankers' effectiveness, provide useful insights with respect to which rankers should be combined. In Ohsumed, WMD shows very low correlation to the other rankers, even though it was the worst effective ranker.

\subsection{Rank Aggregation Results}

We report the rank aggregation results obtained by our method and by the baselines, for each dataset and each of the three combinations of rankers.
From the evaluation procedure previously presented, the following combinations of rankers were selected per dataset:
\begin{itemize}
\item Brodatz: all $3$ rankers; LAS + CCOM; LAS + LBP.
\item Soccer: all $3$; BIC + ACC; BIC + GCH.
\item MPEG-7: all $6$; ASC + AIR; AIR + CFD.
\item Ohsumed: all $7$; BoW-cosine + 2grams-cosine; BoW-cosine + WMD.
\item UKBench: all $7$; VOC + ACC; VOC + ACC + CNN-Caffe.
\item UW: all $12$; JAC + BIC; JAC + OKAPI.
\end{itemize}
Recall that the use of LAS + LBP and BIC + GCH for the Brodatz and the Soccer datasets, respectively, were defined according to Equation~\ref{eq:rankerPairSelection}. The same approach was used for the other datasets.
For UKBench, both second and third selection approaches lead to the same pair of rankers, so we also present the aggregation using its three most effective rankers.

\begin{figure}[htb]
\centering
\includegraphics[width=0.7\linewidth]{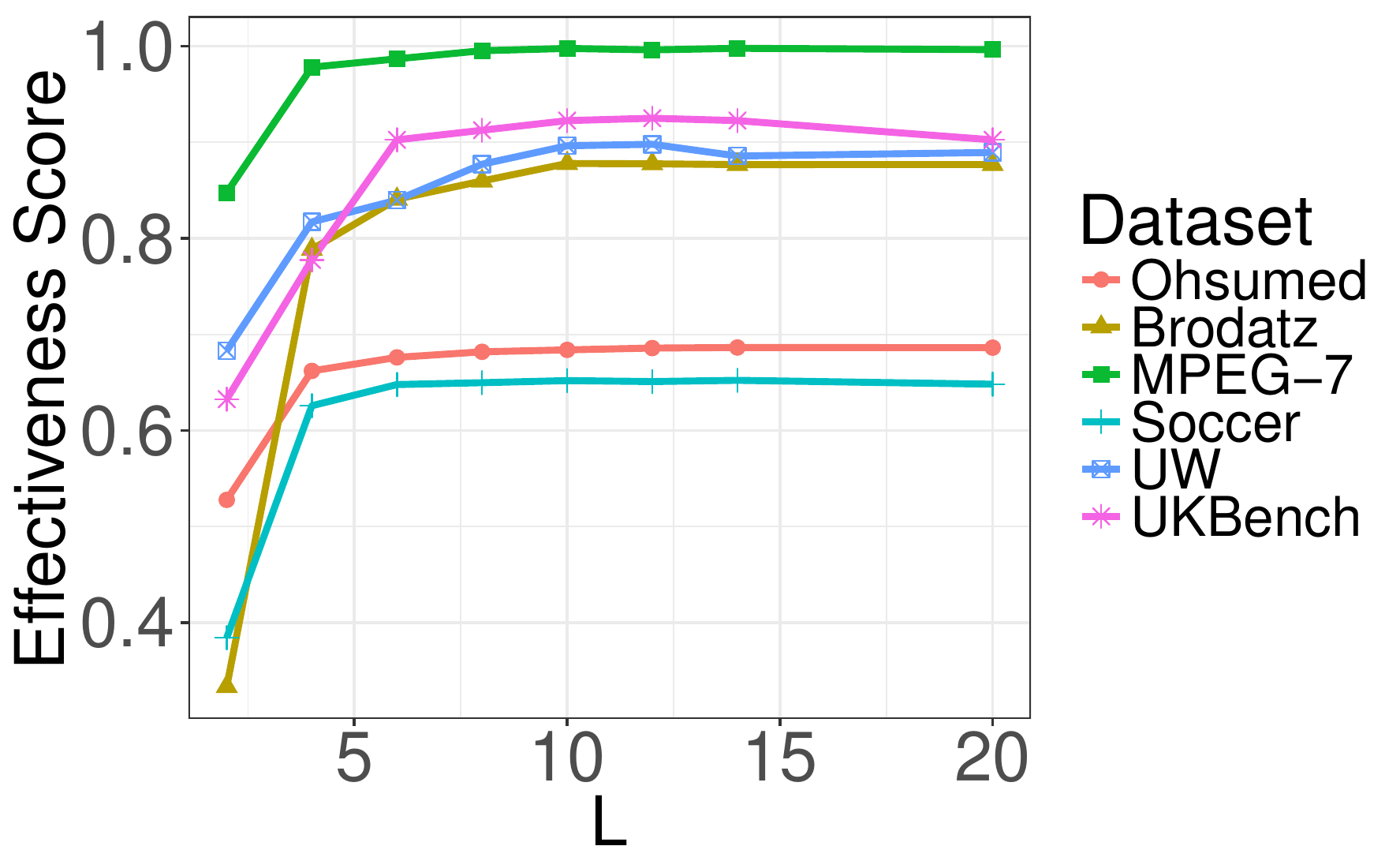}
\caption{\changed{Effect of the cut-off parameter $L$ in the effectiveness performance. N-S scores for UKBench are rescaled to $[0, 1]$ for consistency in the plot, while the rest corresponds to NDCG@10.}}\label{fig:effectParameterL}
\end{figure}

\changed{We performed experiments for different values of $L$ in the range $\{2, 4, 6, 8, 10, 12, 14, 16, 20\}$ for different datasets. The experimental results, using the fusion of all selected rankers for each dataset is presented in Figure~\ref{fig:effectParameterL}. As we can observed, the effectiveness increased when $L$ varies from 2 to 10, and from that point on, the effectiveness measure roughly stabilized.}

\changed{In order to evaluate the impact of different fusion graph comparators in our method, we present in Table~\ref{tab:effectGraphDis} the effectiveness scores achieved by either WGU or MCS, measured by N-S for UKBench, and NDCG@10 for the rest.
The performance is slightly better with WGU than with MCS in absolute values, but WGU was also statistically superior in two out of six cases.
The evaluation comprised all datasets, using the fusion of all selected rankers for them.
Therefore, in the remaining experiments performed, WGU is adopted.}

\begin{table}[ht]
\centering
\caption{\changed{Effect of different fusion graph comparators in the effectiveness performance.}}
\label{tab:effectGraphDis}
\begin{tabular}{lrrrc}
\hline
Collection & \multicolumn{2}{c}{Effectiveness} & Difference (p.p.) & Statistical Difference \\
 & WGU & MCS & & \\\hline
Ohsumed & 0.683835 & 0.677880 & 0.5955 & $\blacktriangle$ \\
Brodatz & 0.878995 & 0.878675 & 0.0320 & $\bullet$ \\
MPEG-7 & 0.997658 & 0.997821 & -0.0163 & $\bullet$ \\
Soccer & 0.651828 & 0.651982 & -0.0154 & $\bullet$ \\
UW & 0.873607 & 0.874543 & -0.0936 & $\bullet$ \\
UKBench & 3.69  & 3.67 & 0.5 & $\blacktriangle$ \\
\hline
\end{tabular}
\end{table}

Tables~\ref{tab:aggResultsOhsumed},
\ref{tab:aggResultsBrodatz},
\ref{tab:aggResultsMPEG7},
\ref{tab:aggResultsSoccer},
\ref{tab:aggResultsUW}, and
\ref{tab:aggResultsUKBench} report the results obtained, respectively for Ohsumed, Brodatz, MPEG-7, Soccer, UW, and UKBench.

\begin{table}
\centering
\caption{Results for rank aggregation on Ohsumed.}
\label{tab:aggResultsOhsumed}
\begin{tabular}{lR{4.7cm}R{1.7cm}R{1.7cm}}\hline
Method & \multicolumn{3}{c}{NDCG@10} \\\hline
 & BoW-cosine + BoW-Jaccard + 2grams-cosine + 2grams-Jaccard + GNF-MCS + GNF-WGU + WMD & BoW-cosine + 2grams-cosine & BoW-cosine + WMD \\\hline
\textbf{FG} & 0.683835 & 0.683472 & 0.676760 \\
RecKNNGraphCCs & $\blacktriangle$ 0.676234 & $\blacktriangle$ 0.679728 & $\blacktriangle$ 0.667750 \\
CombSUM & $\blacktriangle$ 0.666997 & $\blacktriangle$ 0.671868 & $\blacktriangle$ 0.598441 \\
CombMED & $\blacktriangle$ 0.666997 & $\blacktriangle$ 0.671868 & $\blacktriangle$ 0.598441 \\
CombMNZ & $\blacktriangle$ 0.666929 & $\blacktriangle$ 0.671869 & $\blacktriangle$ 0.598261 \\
QueryRankFusion & $\blacktriangle$ 0.651279 & $\blacktriangle$ 0.671258 & $\blacktriangle$ 0.669704 \\
MRA & $\blacktriangle$ 0.666045 & $\blacktriangle$ 0.670357 & $\blacktriangle$ 0.582049 \\
RRF & $\blacktriangle$ 0.665793 & $\blacktriangle$ 0.671294 & $\blacktriangle$ 0.571016 \\
BordaCount & $\blacktriangle$ 0.660197 & $\blacktriangle$ 0.671147 & $\blacktriangle$ 0.570466 \\
Condorcet & $\blacktriangle$ 0.619869 & $\blacktriangle$ 0.670235 & $\blacktriangle$ 0.569906 \\
CombMAX & $\blacktriangle$ 0.611777 & $\blacktriangle$ 0.671305 & $\blacktriangle$ 0.597080 \\
CombANZ & $\blacktriangle$ 0.567671 & $\blacktriangle$ 0.670550 & $\blacktriangle$ 0.595983 \\
Kemeny & $\blacktriangle$ 0.543564 & $\blacktriangle$ 0.665817 & $\blacktriangle$ 0.526588 \\
CombMIN & $\blacktriangle$ 0.502482 & $\blacktriangle$ 0.666559 & $\blacktriangle$ 0.591361 \\
RLSim & $\blacktriangle$ 0.434614 & $\blacktriangle$ 0.639004 & $\blacktriangle$ 0.579972 \\
CorGraph & $\blacktriangle$ 0.487177 & $\blacktriangle$ 0.497431 & $\blacktriangle$ 0.456434 \\
RkGraph & $\blacktriangle$ 0.289045 & $\blacktriangledown$ 0.688443 & $\blacktriangle$ 0.288436 \\
\hline
\end{tabular}
\end{table}

\begin{table}
\centering
\caption{Results for rank aggregation on Brodatz.}
\label{tab:aggResultsBrodatz}
\resizebox{.99\columnwidth}{!}{\begin{tabular}{lrrr}\hline
Method & \multicolumn{3}{c}{NDCG@10} \\\hline
 & LAS+CCOM+LBP & LAS+CCOM & LAS+LBP \\\hline
RecKNNGraphCCs & $\bullet$ 0.877882 & $\blacktriangledown$ 0.882903 & $\blacktriangledown$ 0.839717 \\
\textbf{FG} & 0.878995 & 0.872084 & 0.835624 \\
RkGraph & $\blacktriangle$ 0.812659 & $\blacktriangle$ 0.861250 & $\blacktriangle$ 0.788682 \\
QueryRankFusion & $\blacktriangle$ 0.850263 & $\blacktriangle$ 0.850438 & $\blacktriangle$ 0.808562 \\
CombMNZ & $\blacktriangle$ 0.822887 & $\blacktriangle$ 0.827517 & $\blacktriangle$ 0.787922 \\
CombSUM & $\blacktriangle$ 0.812971 & $\blacktriangle$ 0.826075 & $\blacktriangle$ 0.784971 \\
CombMED & $\blacktriangle$ 0.812971 & $\blacktriangle$ 0.826075 & $\blacktriangle$ 0.784971 \\
CombMAX & $\blacktriangle$ 0.787828 & $\blacktriangle$ 0.818125 & $\blacktriangle$ 0.776842 \\
RRF & $\blacktriangle$ 0.818656 & $\blacktriangle$ 0.817139 & $\blacktriangle$ 0.788840 \\
BordaCount & $\blacktriangle$ 0.805699 & $\blacktriangle$ 0.814664 & $\blacktriangle$ 0.785836 \\
MRA & $\blacktriangle$ 0.822778 & $\blacktriangle$ 0.813396 & $\blacktriangle$ 0.788883 \\
CombANZ & $\blacktriangle$ 0.763987 & $\blacktriangle$ 0.812431 & $\blacktriangle$ 0.769743 \\
Condorcet & $\blacktriangle$ 0.781129 & $\blacktriangle$ 0.809929 & $\blacktriangle$ 0.781781 \\
CorGraph & $\blacktriangle$ 0.749420 & $\blacktriangledown$ 0.895623 & $\blacktriangle$ 0.719204 \\
CombMIN & $\blacktriangle$ 0.713228 & $\blacktriangle$ 0.794631 & $\blacktriangle$ 0.752268 \\
Kemeny & $\blacktriangle$ 0.719680 & $\blacktriangle$ 0.786537 & $\blacktriangle$ 0.757349 \\
RLSim & $\blacktriangle$ 0.633157 & $\blacktriangle$ 0.756053 & $\blacktriangle$ 0.724879 \\\hline
\end{tabular}}
\end{table}

\begin{table}
\centering
\caption{Results for rank aggregation on MPEG-7.}
\label{tab:aggResultsMPEG7}
\begin{tabular}{lR{3.3cm}rr}\hline
Method & \multicolumn{3}{c}{NDCG@10} \\\hline
 & AIR + CFD + ASC + IDSC + BAS + SS
 & ASC + AIR
 & AIR + CFD \\\hline
\textbf{FG} & 0.997658 & 0.994729 & 0.995886 \\
RecKNNGraphCCs & $\bullet$ 0.998052 & $\bullet$ 0.995160 & $\bullet$ 0.997267 \\
RkGraph & $\blacktriangle$ 0.826119 & $\blacktriangledown$ 0.999350 & $\blacktriangle$ 0.992078 \\
CorGraph & $\blacktriangle$ 0.992456 & $\blacktriangle$ 0.962951 & $\blacktriangle$ 0.961460 \\
RRF & $\blacktriangle$ 0.980638 & $\blacktriangle$ 0.957684 & $\blacktriangle$ 0.954499 \\
MRA & $\blacktriangle$ 0.980086 & $\blacktriangle$ 0.950442 & $\blacktriangle$ 0.946144 \\
CombMNZ & $\blacktriangle$ 0.976832 & $\blacktriangle$ 0.942705 & $\blacktriangle$ 0.932234 \\
BordaCount & $\blacktriangle$ 0.974697 & $\blacktriangle$ 0.954296 & $\blacktriangle$ 0.951316 \\
CombSUM & $\blacktriangle$ 0.969212 & $\blacktriangle$ 0.941585 & $\blacktriangle$ 0.930685 \\
CombMED & $\blacktriangle$ 0.969212 & $\blacktriangle$ 0.941585 & $\blacktriangle$ 0.930685 \\
QueryRankFusion & $\blacktriangle$ 0.940976 & $\blacktriangle$ 0.941762 & $\blacktriangle$ 0.941271 \\
CombMAX & $\blacktriangle$ 0.930012 & $\blacktriangle$ 0.941585 & $\blacktriangle$ 0.930685 \\
Condorcet & $\blacktriangle$ 0.911624 & $\blacktriangle$ 0.950122 & $\blacktriangle$ 0.947416 \\
CombANZ & $\blacktriangle$ 0.862366 & $\blacktriangle$ 0.938649 & $\blacktriangle$ 0.927035 \\
Kemeny & $\blacktriangle$ 0.792694 & $\blacktriangle$ 0.940479 & $\blacktriangle$ 0.929906 \\
CombMIN & $\blacktriangle$ 0.626645 & $\blacktriangle$ 0.902798 & $\blacktriangle$ 0.888723 \\
RLSim & $\blacktriangle$ 0.444817 & $\blacktriangle$ 0.902798 & $\blacktriangle$ 0.888723 \\\hline
\end{tabular}
\end{table}

\begin{table}
\centering
\caption{Results for rank aggregation on Soccer.}
\label{tab:aggResultsSoccer}
\resizebox{.99\columnwidth}{!}{\begin{tabular}{lrrr}\hline
Method & \multicolumn{3}{c}{NDCG@10} \\\hline & BIC+ACC+GCH & BIC+ACC & BIC+GCH \\\hline
\textbf{FG} & 0.651828 & 0.655332 & 0.622217 \\
RkGraph & $\bullet$ 0.653623 & $\bullet$ 0.656422 & $\bullet$ 0.628563 \\
CorGraph & $\blacktriangle$ 0.645004 & $\blacktriangle$ 0.643505 & $\bullet$ 0.623627 \\
RecKNNGraphCCs & $\blacktriangle$ 0.637537 & $\blacktriangle$ 0.640729 & $\bullet$ 0.618704 \\
QueryRankFusion & $\blacktriangle$ 0.613732 & $\blacktriangle$ 0.613659 & $\blacktriangle$ 0.598862 \\
BordaCount & $\blacktriangle$ 0.603156 & $\blacktriangle$ 0.613205 & $\blacktriangle$ 0.589633 \\
RRF & $\blacktriangle$ 0.604119 & $\blacktriangle$ 0.613005 & $\blacktriangle$ 0.590819 \\
CombSUM & $\blacktriangle$ 0.604575 & $\blacktriangle$ 0.611546 & $\blacktriangle$ 0.588667 \\
CombMED & $\blacktriangle$ 0.604565 & $\blacktriangle$ 0.611546 & $\blacktriangle$ 0.588667 \\
CombMNZ & $\blacktriangle$ 0.605567 & $\blacktriangle$ 0.611269 & $\blacktriangle$ 0.589202 \\
MRA & $\blacktriangle$ 0.605971 & $\blacktriangle$ 0.611017 & $\blacktriangle$ 0.588399 \\
CombANZ & $\blacktriangle$ 0.587048 & $\blacktriangle$ 0.610981 & $\blacktriangle$ 0.582010 \\
Condorcet & $\blacktriangle$ 0.593809 & $\blacktriangle$ 0.611049 & $\blacktriangle$ 0.589266 \\
CombMAX & $\blacktriangle$ 0.591911 & $\blacktriangle$ 0.609345 & $\blacktriangle$ 0.584983 \\
Kemeny & $\blacktriangle$ 0.578919 & $\blacktriangle$ 0.607451 & $\blacktriangle$ 0.578043 \\
CombMIN & $\blacktriangle$ 0.570258 & $\blacktriangle$ 0.606144 & $\blacktriangle$ 0.576877 \\
RLSim & $\blacktriangle$ 0.506736 & $\blacktriangle$ 0.570744 & $\blacktriangle$ 0.545591 \\\hline
\end{tabular}}
\end{table}

\begin{table}
\centering
\caption{Results for rank aggregation on UW.}
\label{tab:aggResultsUW}
\begin{tabular}{lR{3.5cm}R{1.7cm}R{1.7cm}}\hline
Method & \multicolumn{3}{c}{NDCG@10} \\\hline
 & JAC + BIC + DICE + BOW + OKAPI + JACCARD + TF-IDF + GCH + COSINE + LAS + HTD + QCCH & JAC + BIC & JAC + OKAPI \\\hline
CorGraph & $\blacktriangledown$ 0.896341 & $\blacktriangle$ 0.842665 & $\blacktriangledown$ 0.933452 \\
\textbf{FG} & 0.873607 & 0.854473 & 0.882776 \\
RecKNNGraphCCs & $\blacktriangle$ 0.869448 & $\blacktriangle$ 0.843423 & $\bullet$ 0.882035 \\
RkGraph & $\blacktriangle$ 0.746804 & $\blacktriangle$ 0.841127 & $\blacktriangle$ 0.866544 \\
MRA & $\blacktriangle$ 0.815983 & $\blacktriangle$ 0.797292 & $\blacktriangle$ 0.786995 \\
RRF & $\blacktriangle$ 0.815779 & $\blacktriangle$ 0.798502 & $\blacktriangle$ 0.795143 \\
CombMNZ & $\blacktriangle$ 0.806416 & $\blacktriangle$ 0.793488 & $\blacktriangle$ 0.814850 \\
BordaCount & $\blacktriangle$ 0.789620 & $\blacktriangle$ 0.797677 & $\blacktriangle$ 0.788127 \\
CombSUM & $\blacktriangle$ 0.769227 & $\blacktriangle$ 0.793057 & $\blacktriangle$ 0.812383 \\
CombMED & $\blacktriangle$ 0.769227 & $\blacktriangle$ 0.793057 & $\blacktriangle$ 0.812383 \\
QueryRankFusion & $\blacktriangle$ 0.747281 & $\blacktriangle$ 0.792681 & $\blacktriangle$ 0.807250 \\
Condorcet & $\blacktriangle$ 0.743168 & $\blacktriangle$ 0.795304 & $\blacktriangle$ 0.780080 \\
CombMAX & $\blacktriangle$ 0.691427 & $\blacktriangle$ 0.786912 & $\blacktriangle$ 0.802788 \\
CombANZ & $\blacktriangle$ 0.596119 & $\blacktriangle$ 0.784183 & $\blacktriangle$ 0.795284 \\
Kemeny & $\blacktriangle$ 0.471099 & $\blacktriangle$ 0.773449 & $\blacktriangle$ 0.739709 \\
CombMIN & $\blacktriangle$ 0.359668 & $\blacktriangle$ 0.773776 & $\blacktriangle$ 0.769389 \\
RLSim & $\blacktriangle$ 0.330593 & $\blacktriangle$ 0.740222 & $\blacktriangle$ 0.768275 \\\hline
\end{tabular}
\end{table}

\begin{table}
\centering
\caption{Results for rank aggregation on UKBench.}
\label{tab:aggResultsUKBench}
\begin{tabular}{lR{4.2cm}R{1.7cm}R{1.7cm}}\hline
Method & \multicolumn{3}{c}{N-S Score} \\\hline
 & VOC + ACC + CNN-Caffe + SCD + JCD + FCTH-SPy + CEDD-SPy & VOC + ACC & VOC + ACC + CNN-Caffe \\\hline
\textbf{FG} & 3.69 & 3.83 & 3.90 \\
RecKNNGraphCCs & $\blacktriangle$ 3.67 & $\blacktriangle$ 3.81 & $\blacktriangle$ 3.87 \\
QueryRankFusion & $\blacktriangle$ 3.60 & $\blacktriangle$ 3.78 & $\blacktriangle$ 3.86 \\
MRA & $\blacktriangle$ 3.52 & $\blacktriangle$ 3.50 & $\blacktriangle$ 3.77 \\
CombSUM & $\blacktriangle$ 3.55 & $\blacktriangle$ 3.60 & $\blacktriangle$ 3.76 \\
CombMED & $\blacktriangle$ 3.55 & $\blacktriangle$ 3.60 & $\blacktriangle$ 3.76 \\
CombMNZ & $\blacktriangle$ 3.53 & $\blacktriangle$ 3.60 & $\blacktriangle$ 3.76 \\
BordaCount & $\blacktriangle$ 3.55 & $\blacktriangle$ 3.60 & $\blacktriangle$ 3.76 \\
RRF & $\blacktriangle$ 3.52 & $\blacktriangle$ 3.60 & $\blacktriangle$ 3.76 \\
Condorcet & $\blacktriangle$ 3.64 & $\blacktriangle$ 3.58 & $\blacktriangle$ 3.75 \\
CombMAX & $\blacktriangle$ 3.13 & $\blacktriangle$ 3.52 & $\blacktriangle$ 3.48 \\
RkGraph & $\blacktriangle$ 3.03 & $\blacktriangle$ 3.50 & $\blacktriangle$ 3.54 \\
CombANZ & $\blacktriangle$ 2.83 & $\blacktriangle$ 3.42 & $\blacktriangle$ 3.28 \\
Kemeny & $\blacktriangle$ 2.51 & $\blacktriangle$ 3.37 & $\blacktriangle$ 3.14 \\
CombMIN & $\blacktriangle$ 2.35 & $\blacktriangle$ 3.36 & $\blacktriangle$ 3.09 \\
CorGraph & $\blacktriangle$ 2.44 & $\blacktriangle$ 2.91 & $\blacktriangle$ 2.77 \\
RLSim & $\blacktriangle$ 1.09 & $\blacktriangle$ 2.73 & $\blacktriangle$ 1.89 \\
\hline
\end{tabular}
\end{table}

Most baselines presented results worse than the best individual rankers, but our method overcame the individual rankers in all scenarios.
It can be seen that most baselines are dramatically affected by bad individual ranks, in the sense that the addition of a poor ranker into the aggregation function leads to poor effectiveness. This may be seen as a negative aspect of unsupervised rank aggregation functions in general. Our method, on the contrary, was shown to be much less sensitive to this search scenario.
For the Ohsumed dataset, for example, WMD performed much worse than BoW-cosine, but, because they produce low correlated ranks, their fusion still yielded a better ranker.

The criteria adopted to choose pairs of rankers for combination, based on effectiveness and correlation, led to pairs whose aggregated results surpassed pairs formed by the most effective rankers for the MPEG-7 and UW datasets. In most cases, the selection of the most effective base rankers yields suitable results.

In Ohsumed, Brodatz, and MPEG-7, the aggregation of all rankers performed better than the combination of selected pairs of rankers. These results demonstrate that even less competitive rankers can contribute to improving retrieval tasks when used in the aggregation.

\changed{While the ranker selection criteria of using all rankers led to top performance in half the datasets, it also demands additional processing cost.
The analysis of the three ranker selection approaches allows us to conclude that the ranker selection of the two most competitive rankers per dataset is an overall good choice, but subjected to improvement after a careful empirical evaluation of the other approaches in the desired scenario.}

We summarize in Table~\ref{tab:globalComparisonsRankers} the results achieved by the rankers in each dataset, and report our gains over them, in percentage gain. \textit{FG} was able to present significant gains over the rankers.

\begin{table*}
\caption{Effectiveness of rankers compared to our method, in textual, image, and hybrid datasets.}
\label{tab:globalComparisonsRankers}
\begin{subtable}[t]{0.32\textwidth}
\centering
\caption{Brodatz}
\label{tab:rankersResultsVsFG_Brodatz}
\resizebox{3.9cm}{!}{%
\begin{tabular}{lrr}\hline
Method & NDCG@10 & Gains (\%) \\\hline
\textbf{FG} & 0.878995 & \\
LAS        & 0.850533 & 3.35 \\
CCOM       & 0.726186 & 21.04 \\
LBP        & 0.652759 & 33.66 \\\hline
\end{tabular}}
\vspace{2mm}
\caption{UW dataset}
\label{tab:rankersResultsVsFG_UW}
\resizebox{3.9cm}{!}{%
\begin{tabular}{lrr}\hline
Method & NDCG@10 & Gains (\%) \\\hline
\textbf{FG} & 0.873607 & \\
JAC & 0.810729 & 7.76 \\
BIC & 0.746454 & 17.03 \\
DICE & 0.722831 & 20.86 \\
BOW & 0.720781 & 21.20 \\
OKAPI & 0.716035 & 22.01 \\
JACCARD & 0.701651 & 24.51 \\
TF-IDF & 0.658880 & 32.59 \\
GCH & 0.630315 & 38.60 \\
COSINE & 0.554767 & 57.47 \\
LAS & 0.514314 & 69.86 \\
HTD & 0.495002 & 76.49 \\
QCCH & 0.414249 & 110.89 \\
\hline
\end{tabular}}
\end{subtable}
\begin{subtable}[t]{0.32\textwidth}
\centering
\caption{MPEG-7}
\label{tab:rankersResultsVsFG_MPEG7}
\resizebox{3.9cm}{!}{%
\begin{tabular}{lrr}\hline
Method & NDCG@10 & Gains (\%) \\\hline
\textbf{FG} & 0.997658 & \\
ASC  & 0.941585 & 5.96 \\
AIR  & 0.939424 & 6.20 \\
CFD  & 0.930685 & 7.20 \\
IDSC & 0.922828 & 8.11 \\
BAS  & 0.866098 & 15.19 \\
SS   & 0.611481 & 63.15 \\
\hline
\end{tabular}}
\vspace{2mm}
\caption{Ohsumed}
\label{tab:rankersResultsVsFG_Ohsumed}
\resizebox{3.9cm}{!}{%
\begin{tabular}{lrr}\hline
Method & NDCG@10 & Gains (\%) \\\hline
\textbf{FG} & 0.683835 & \\
BoW-cosine & 0.669701 & 2.11 \\
2grams-cosine & 0.664120 & 2.97 \\
GNF-WGU & 0.662668 & 3.19 \\
GNF-MCS & 0.655420 & 4.34 \\
2grams-Jaccard & 0.651320 & 4.99 \\
BoW-Jaccard & 0.645711 & 5.90 \\
WMD & 0.427361 & 60.01 \\
\hline
\end{tabular}}
\end{subtable}
\begin{subtable}[t]{0.32\textwidth}
\centering
\caption{UKBench}
\label{tab:rankersResultsVsFG_UKBench}
\resizebox{3.9cm}{!}{%
\begin{tabular}{lrr}\hline
Method & N-S Score & Gains (\%) \\\hline
\textbf{FG} & 3.90 & \\
VOC & 3.54 & 10.17 \\
ACC & 3.37 & 15.73 \\
CNN-Caffe & 3.31 & 17.83 \\
SCD & 3.15 & 23.81 \\
JCD & 2.79 & 39.79 \\
FCTH-SPy & 2.73 & 42.86 \\
CEDD-SPy & 2.61 & 49.43 \\
\hline
\end{tabular}}
\vspace{2mm}
\caption{Soccer}
\label{tab:rankersResultsVsFG_Soccer}
\resizebox{3.9cm}{!}{%
\begin{tabular}{lrr}\hline
Method & NDCG@10 & Gains (\%) \\\hline
\textbf{FG} & 0.655332 & \\
BIC & 0.614818 & 6.59 \\
ACC & 0.592699 & 10.57 \\
GCH & 0.536412 & 22.17 \\
\hline
\end{tabular}}
\end{subtable}
\end{table*}


Our method achieved either top or very competitive performance in all datasets and combinations of rankers tested. For $6$ datasets with $3$ configurations each, and for $16$ state-of-the-art baselines, \textit{FG} was statistically worse only in $7$ out of $288$ comparisons. Besides, it was top-$1$ in $4$ out of $6$ datasets (Ohsumed, MPEG-7, Soccer, and UKBench), and top-$2$ in Brodatz and UW.

\begin{figure}[ht]
\includegraphics[width=\linewidth]{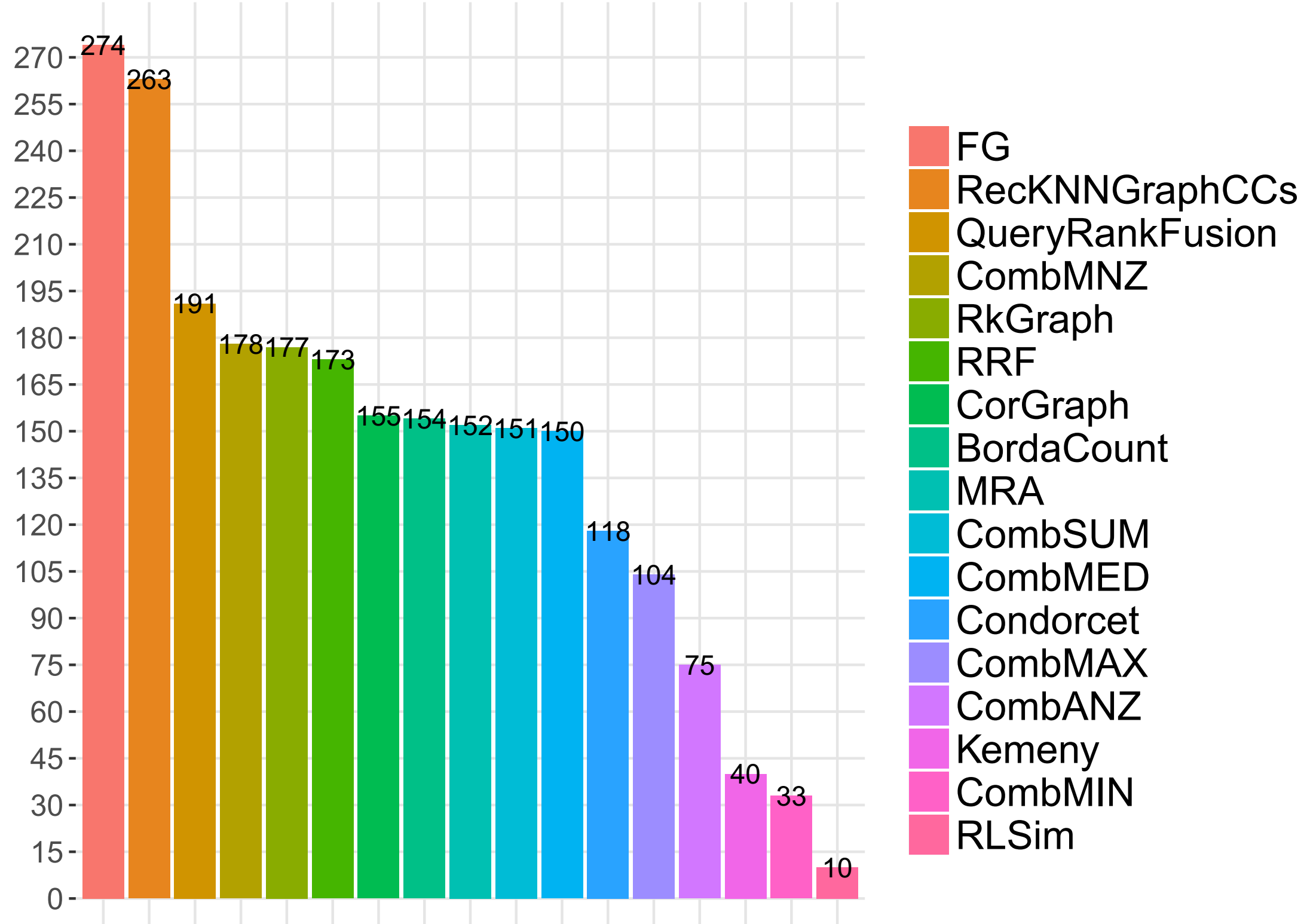}
\caption{Winning numbers achieved per rank aggregation function.}\label{fig:winningNumbers}
\end{figure}

We present in Figure~\ref{fig:winningNumbers} the winning numbers achieved per rank aggregation function, in order to contrast them globally. \textit{FG} was broadly superior to the majority of baselines, according to our experimental evaluation comprising $3$ aggregation approaches for each of the $6$ distinct datasets used.

Table~\ref{tab:artOfTheArtUKBench} presents the results of our ranker for UKBench, together with seven additional baselines.
The table reports the results from Table~\ref{tab:aggResultsUKBench}, obtained using ACC + VOC + CNN-Caffe, together with results reported by the other baselines.
QueryRankFusion is presented twice, one regarding their own reported result, and another considering the same input rankers as ours. It is worth to notice that three of these additional baselines performed worse than some classic and much simpler rank aggregation functions. Again, our method achieved the best performance.

\begin{table}
\centering
\caption{State-of-the-art results on UKBench. Results marked with * were obtained using ACC + VOC + CNN-Caffe.}
\label{tab:artOfTheArtUKBench}
\begin{tabular}{ll}\hline
Method & N-S Score \\
\hline
\textbf{FG} & \textbf{3.90*} \\
\citet{xie:2015:image} & 3.89 \\
RecKNNGraphCCs & 3.87* \\
\citet{bai:2016:sparse} & 3.86 \\
QueryRankFusion & 3.86* \\
\citet{zheng:2015:query} & 3.84 \\
QueryRankFusion & 3.83 \\
MRA & 3.77* \\
CombSUM & 3.76* \\
CombMED & 3.76* \\
CombMNZ & 3.76* \\
BordaCount & 3.76* \\
RRF & 3.76* \\
Condorcet & 3.75* \\
\citet{wang:2012:metricfusion} & 3.68 \\
\citet{qin:2011:objRetrieval} & 3.67 \\
\citet{zheng:2014:lpnorm} & 3.57 \\
RkGraph & 3.54* \\
CombMAX & 3.48* \\
CombANZ & 3.28* \\
Kemeny & 3.14* \\
CombMIN & 3.09* \\
CorGraph & 2.91* \\
RLSim & 1.89* \\
\hline
\end{tabular}
\end{table}

Our rank fusion has been shown to be effective in combining contextual information from different ranks, along with the intrinsic relationships that the retrieved objects have to each other in their own ranks. Also, our procedure to rank objects based on fusion graphs considers these fusions automatically, relying on such graphs without any other intermediate steps, such as trainning or parameter tuning.

\subsection{Efficiency Analysis}

Section~\ref{computationalAnalysis} presents the asymptotic cost of our method. The time for performing a query is around the sum of the slowest time to produce an isolated rank plus the time to produce the final rank based on fusion graphs.
\changed{Table~\ref{tab:costRankAgg} presents, per dataset, the mean time (in milliseconds) spent per query, and the mean offline time (in seconds) to produce the fusion graphs from the response set. For both values, we report the mean time of $5$ independent measurements, taken on an Intel Core i7-7500U CPU @ 2.70GHz with 16GB of RAM. 
For all datasets, the search time was reasonable, given the high gains in effectiveness provided by our method. The offline time is also low, due to unsupervised nature of our method.}

\begin{table}[ht]
\centering
\caption{\changed{Rank aggregation time per query, and offline time.}}
\label{tab:costRankAgg}
\begin{tabular}{lrr}\hline
Dataset & Rank Aggregation Time (in ms) & Offline Time (in sec) \\\hline
Brodatz & $8.76 \pm 2.70$ & $2.28 \pm 0.75$ \\
MPEG-7 & $25.73 \pm 2.76$ & $5.92 \pm 0.60$ \\
Ohsumed & $101.13 \pm 16.88$ & $38.12 \pm 3.78$ \\
Soccer & $4.60 \pm 2.58$ & $0.63 \pm 0.16$ \\
UW & $29.29 \pm 7.80$ & $10.92 \pm 1.48$ \\
UKBench & $21.30 \pm 6.68$ & $25.34 \pm 2.02$ \\
\hline
\end{tabular}
\end{table}

\section{Conclusions}
\label{secConc}

Distinct features provide different and complementary views of textual and multimedia documents in retrieval tasks.
Therefore, combining such results for a more effective retrieval without the need of user intervention remains a relevant and challenging task.

In this paper, a novel unsupervised graph-based rank aggregation method is proposed. Our approach models the rank fusion task by means of a fusion graph and derives a novel fused retrieval score, directly based on the graph structure. The method was extensively evaluated on textual, images, and hybrid datasets \changed{comprising ad-hoc retrieval tasks}, achieving superior effectiveness scores than the best isolated features and several baselines.

As a future work, we intend to evaluate our method against supervised techniques. We also want to explore other rank-fusion vector representations based on graphs. The goal is to take advantages of existing solutions (e.g., indexing schemes) to make our fusion method even more scalable.


\section*{References}

\bibliography{bib}

\end{document}